\documentclass[journal,draftclsnofoot,onecolumn,12pt,twoside]{IEEEtran}
\usepackage[T1]{fontenc}% optional T1 font encoding
\usepackage{cite}
\usepackage{amsmath,amssymb,amsfonts}
\usepackage{algorithm}
\usepackage{algorithmic}
\usepackage{graphicx}
\usepackage{textcomp}
\usepackage{xcolor}
\usepackage{subfigure}
\usepackage{booktabs}
\usepackage{multirow}
\usepackage{url}
\usepackage{threeparttable}
\usepackage{array}
\usepackage{textcase}
\usepackage{rotating}
\usepackage{etoolbox}

\graphicspath{{figures/}}

\def\nt{N_{\rm t}}
\def\nr{N_{\rm r}}

\begin{document}

\ifdefined \GramaCheck
  \newcommand{\CheckRmv}[1]{}
  \newcommand{\figref}[1]{Figure 1}%
  \newcommand{\tabref}[1]{Table 1}%
  \newcommand{\secref}[1]{Section 1}
  \newcommand{\algref}[1]{Algorithm 1}
  \renewcommand{\eqref}[1]{Equation 1}
\else
  \newcommand{\CheckRmv}[1]{#1}
  \newcommand{\figref}[1]{Fig.~\ref{#1}}%
  \newcommand{\tabref}[1]{Table~\ref{#1}}%
  \newcommand{\secref}[1]{Sec.~\ref{#1}}
  \newcommand{\algref}[1]{Algorithm~\ref{#1}}
  \renewcommand{\eqref}[1]{(\ref{#1})}
\fi

\title{Conditional Diffusion Model-Enabled Scenario-Specific Neural Receivers for Superimposed Pilot Schemes} 
\author{Xingyu~Zhou,
        Le~Liang,
		    Xinjie~Li,	
		    Jing~Zhang,\\
		    Peiwen~Jiang,
        Xiao~Li,
        and~Shi~Jin% <-this % stops a space
\thanks{Xingyu Zhou, Le~Liang, Xinjie~Li, Jing Zhang, Peiwen~Jiang, Xiao Li, and Shi Jin are with the School of Information Science and Engineering, Southeast University, Nanjing 210096, China
(e-mail: \protect \url{xy_zhou@seu.edu.cn}; lliang@seu.edu.cn; lixinjie@seu.edu.cn; jingzhang@seu.edu.cn; peiwenjiang@seu.edu.cn; \protect \url{li_xiao@seu.edu.cn}; jinshi@seu.edu.cn). Le~Liang is also with Purple Mountain Laboratories, Nanjing 211111, China. \textit{(Corresponding author: Shi Jin; Le Liang.)}}}

\maketitle

\begin{abstract}
Neural receivers have demonstrated strong performance in wireless communication systems. However, their effectiveness typically depends on access to large-scale, scenario-specific channel data for training, which is often difficult to obtain in practice. Recently, generative artificial intelligence (AI) models, particularly diffusion models (DMs), have emerged as effective tools for synthesizing high-dimensional data. This paper presents a scenario-specific channel generation method based on conditional DMs, which accurately model channel distributions conditioned on user location and velocity information. The generated synthetic channel data are then employed for data augmentation to improve the training of a neural receiver designed for superimposed pilot-based transmission. Experimental results show that the proposed method generates high-fidelity channel samples and significantly enhances neural receiver performance in the target scenarios, outperforming conventional data augmentation and generative adversarial network-based techniques.
\end{abstract}


%%%%%%%%%%%%%%%%%%%%%%%%%%%%%%%%%%%%%%%%%%%%%%%%%%%%%%%%%%
\section{Introduction}  % 连摘要2 page
\label{sec:intro}

% -- background: 需求(MIMO->accurate CSI->pilot overhead)-->sip transmission-->receiver重要性
Channel state information (CSI) is essential for fully exploiting the benefits of multiple-input multiple-output (MIMO) technology in wireless communication systems. However, acquiring accurate CSI typically requires a large number of pilot symbols, which are time-multiplexed with data symbols and result in substantial transmission overhead. To improve spectral efficiency, superimposed pilot (SIP) schemes have attracted increasing interest \cite{verenzuela2018spectral,gan2024bayesian,xie2024superimposed}. In these schemes, pilot symbols are embedded within data symbols instead of being allocated to dedicated time-frequency resources, as in conventional orthogonal pilot (OP) transmission. While SIP schemes reduce pilot overhead, especially in high-mobility environments, they introduce challenges such as mutual interference between pilot and data symbols, which complicates channel estimation and symbol detection tasks \cite{qian2024enhancing}. 

% neural receiver-->  limitation(massive MIMO high-dimensional channel, data hard to get)
In recent years, artificial intelligence (AI) and deep learning (DL) have shown great potential in solving problems that are difficult to address through conventional wireless system design approaches \cite{qin2024ai}. In particular, neural receivers, DL-based architectures that learn to demodulate signals directly from raw input data, have emerged as promising solutions for overcoming the limitations of SIP schemes. Their ability to capture nonlinear relationships and adapt to dynamic channel conditions makes them especially useful in such settings \cite{xiao2025interference,gu2024learning,li2025ai,ait2021end}. However, the effectiveness of neural receivers heavily depends on the availability of large-scale, diverse training datasets that accurately reflect realistic channel conditions. Acquiring such datasets remains a significant challenge due to limitations in channel measurement, storage, and computational resources \cite{lee2024generating}. This issue is further exacerbated as wireless systems evolve toward ultra-massive MIMO architectures, which involve extremely high-dimensional channel representations \cite{wang2024tutorial}. Additionally, channel characteristics vary rapidly across different environments, making it difficult for neural receivers to generalize effectively. The scenario-adaptability of these models therefore remains a critical challenge. This has motivated growing interest in data augmentation techniques, which aim to synthetically enrich training datasets and improve the robustness and generalization capability of neural receivers.

Recent developments in generative AI have introduced transformative opportunities in wireless communications \cite{bariah2024large}. Deep generative models have been successfully applied in tasks such as channel modeling \cite{xiao2022channelgan,sengupta2023generative}, channel estimation \cite{fesl2024diffusion,zhou2025generative}, and data augmentation for downstream learning tasks \cite{chi2024rf,baur2024evaluation}. Among these models, diffusion models (DMs) \cite{sohl2015deep,ho2020denoising,dhariwal2021diffusion} have demonstrated exceptional performance in generating high-fidelity data and capturing complex data distributions. Despite their potential, the use of DMs for scenario-specific data generation and augmentation in learning-based communication systems remains relatively unexplored. In particular, their ability to support neural receiver training through environment-aware synthetic channel generation has not been fully investigated.  

A key advantage of DMs lies in their ability to generate conditional data distributions, making them suitable for capturing the relationships between environmental contexts, such as user location and mobility, and channel behavior. This conditional generation capability enables tailored, scenario-specific data synthesis. A prior study \cite{lee2024generating} proposed a DM-based approach to generate channel samples based on user equipment (UE) location. However, this method did not account for the coexistence of diverse environmental conditions within a typical urban cell \cite{li2023multi}, and it lacked mechanisms for enhancing neural receiver performance in specific scenarios.

In this paper, we develop a DM-based channel data generation framework aimed at scenario-specific augmentation of neural receivers in SIP-based transmissions. By conditioning the model on environmental inputs, including UE location and speed, the proposed DM effectively captures key channel features such as angular, delay, and temporal characteristics. The model is trained using a limited amount of high-fidelity channel data generated from simulation tools. A consistency distillation strategy is further incorporated to significantly reduce the generation complexity and latency, enabling practical deployment. The synthetic channel samples closely match the true distribution of the target scenario and are used to construct an augmented training dataset for the neural receiver. Simulation results show that the proposed approach significantly improves the performance of neural receivers under SIP schemes, outperforming conventional data augmentation techniques and generative adversarial network (GAN)-based methods. Its performance approaches that achieved by models trained with abundant true channel data.

The main contributions of this paper are summarized as follows:

\begin{itemize}
\item \textbf{Conditional DM Framework for Scenario-Specific Channel Generation:} We introduce a DM-based framework that generates scenario-specific channel samples conditioned on UE location and speed. This method outperforms GAN-based alternatives in distributional similarity metrics and significantly enhances neural receiver performance in the targeted scenarios.

\item \textbf{Dual U-Net Architecture for 3D CSI Feature Learning:} A specialized DM architecture is developed using two sequential U-Nets that operate in distinct two-dimensional (2D) domains of the three-dimensional (3D) CSI, effectively learning angular, delay, and temporal characteristics of the channel.

\item \textbf{Consistency Distillation for Efficient Inference:} To address the high computational cost of standard DMs, we employ a consistency distillation scheme that accelerates inference by a factor of 80 while maintaining the fidelity of generated samples, enabling practical real-time application.
\end{itemize}

\textit{Notations:} Bold uppercase and lowercase letters denote matrices and vectors, respectively. $(\cdot)^{\rm H}$ is the Hermitian transpose, and $\odot$ denotes the Hadamard product. $\mathbf{0}$ and $\mathbf{I}$ represent the zero vector and identity matrix, while $\|\cdot\|_2$ is the $l_2$ norm. $\mathbb{E}[\cdot]$ denotes expectation. $\mathcal{N}(z;\mu, \sigma^2)$ and $\mathcal{U}(\mathcal{A})$ denote Gaussian and uniform distributions, respectively. $\mathbb{R}$ and $\mathbb{C}$ represent the sets of real and complex numbers, respectively.

% For discrete sets, $[N]$ represents nonnegative integers up to $N$, and $\mathcal{U}([N])$ denotes a uniform distribution over $[N]$.

%%%%%%%%%%%%%%%%%%%%%%%%%%%%%%%%%%%%%%%%%%%%%%%%%%%%%%%%%%
\section{Problem Formulation}  % 1.5 page
\label{sec:problem}

\subsection{System Model}
% 只考虑单用户的上行接收, 
% -- channel model 
Consider a point-to-point uplink MIMO system in which the base station (BS) is equipped with $\nr$ receive antennas, and the UE  is equipped with $\nt$ transmit antennas. Both the BS and UE employ uniform linear arrays (ULAs) with half-wavelength spacing between antenna elements. 
% time duration and number of subcarriers, resource blocks  (信道已涉及time, frequency-domain)
An orthogonal frequency division multiplexing (OFDM) scheme is adopted, where each frame consists of $K$ subcarriers and $S$ consecutive symbols, resulting in $V = KS$ time-frequency resource elements (REs). The channel is modeled using a geometric channel model. The complex-valued channel matrix for the $(k, s)$-th RE, denoted by $\mathbf{H}_{\rm cplx}^{(k, s)} \in \mathbb{C}^{\nr \times \nt}$ for $k \in \{1, \ldots, K\}$ and $s \in \{1, \ldots, S\}$, is expressed as % \footnote{Simulated using QuaDRiGa according to the geometry-based stochastic channel models --- \cite{li2023multi}}
\CheckRmv{
  \begin{equation}
    \mathbf{H}_{\rm cplx}^{(k, s)} =\sum_{l=1}^{L}\gamma_{l}^{(k, s)} \mathbf{a}_{\rm r}\left(\theta_{l}\right)\mathbf{a}_{\rm t}^{\rm H}(\phi_{l}),  %\;\;{\rl \theta_i^{(s)}, \phi_i^{(s)}}  % 参考Dai channel prediction文章，角度不时变
    \label{eq:channel}
  \end{equation}
}
where $L$ denotes the number of propagation paths, and $\gamma_{l}^{(k, s)}$ represents the complex gain of the $l$-th path at the $(k, s)$-th RE. The vectors $\mathbf{a}_{\rm t}(\cdot)$ and $\mathbf{a}_{\rm r}(\cdot)$ denote the transmit and receive array response vectors, respectively. The parameters $\phi_l$ and $\theta_l$ represent the azimuth angles of departure and arrival for the $l$-th path. By stacking $\mathbf{H}_{\rm cplx}^{(k, s)}$ across the frequency and time dimensions, the complete channel tensor is obtained as $\mathbf{H}_{\rm cplx}\in\mathbb{C}^{\nr \times \nt \times K \times S}$.

% -- Signal model: SIP transmission(编解码等不讲, 对DA和receiver介绍好讲，主体参照OPPO文章走) 
% 先总说SIP特点：pilot and data symbols are superimposed on each RE; 讲优势
The UE employs SIP transmission, where the transmitted block from the $n$-th transmit antenna ($n \in \{1, \ldots, \nt\}$) is constructed by superimposing a pilot matrix $\mathbf{P}_n \in \mathbb{C}^{K \times S}$ and a data matrix $\mathbf{D}_n \in \mathbb{C}^{K \times S}$ as follows 
\CheckRmv{
  \begin{equation}
    \mathbf{S}_n = \sqrt{\rho}\mathbf{P}_n  + \sqrt{1-\rho} \mathbf{D}_n,
  \end{equation}
}
where $\rho \in [0,1]$ denotes the power allocation factor between pilot and data signals. The received signal at the $m$-th antenna of the BS, denoted as $\mathbf{Y}_m \in \mathbb{C}^{K \times S}$, can be expressed as
\CheckRmv{
  \begin{equation}
    \mathbf{Y}_m = \sum_{n=1}^{\nt} \mathbf{H}_{{\rm cplx}, m,n} \odot \mathbf{S}_n + \mathbf{N}_m,
    \label{eq:receive}
  \end{equation}
}
where $\mathbf{H}_{{\rm cplx}, m,n}\in \mathbb{C}^{K\times S}$ is the frequency-time domain channel matrix between the $n$-th transmit antenna and the $m$-th receive antenna, extracted as a slice from the full channel tensor $\mathbf{H}_{{\rm cplx}}$. The matrix $\mathbf{N}_m$ represents additive white Gaussian noise (AWGN).

% Task/Problem formulation: SIP receiver; Challenge: 1) interference 2) requirement of channel data
The goal of the receiver is to recover the transmitted data symbols and decode the corresponding message bits. Conventional approaches typically rely on separate modules for channel estimation and symbol detection. However, SIP transmission introduces mutual interference between pilot and data components, which complicates the separation of the two signals and increases the difficulty of accurate recovery. To address this issue, the neural receiver proposed in \cite{xiao2025interference} leverages data-driven neural networks (NNs) to jointly perform channel estimation and data detection, while incorporating an interference cancellation mechanism. Despite its effectiveness, this approach requires a large-scale training dataset with approximately $10^6$ channel samples to achieve reliable accuracy and generalization. Such a requirement imposes considerable burdens in terms of channel measurement, data collection, and storage.

\subsection{Channels Across Multiple Scenarios} 
\CheckRmv{
  \begin{figure*}[t]
    \centering
    \includegraphics[width=6in]{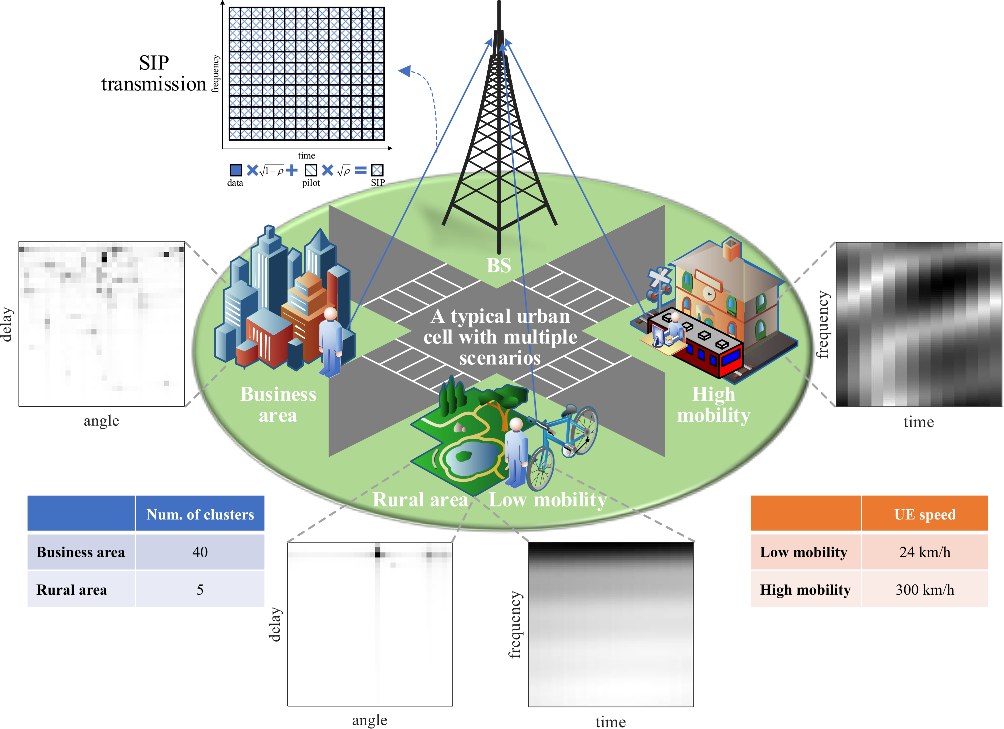}
    \caption{A typical urban cell containing multiple scenarios, where the uplink transmission adopts the SIP scheme. The channel sample of each scenario is displayed using grayscale map.}
    \label{fig:scenario}
  \end{figure*}
}

% 结构： multiple scenarios -- scattering environments, different number of clusters -- multipath in angular-delay domain, different angle and delay spread and range.
An urban micro-cell environment typically comprises multiple scenarios with distinct characteristics \cite{li2023multi}, which are reflected in the UE's surrounding scattering conditions and mobility patterns, as illustrated in \figref{fig:scenario}. These diverse scenarios result in significantly different channel distributions, which in turn influence signal propagation and receiver design. 

We begin by examining the impact of the UE's surrounding scatterers, which are primarily determined by its location, assuming a fixed environmental configuration within the cell. For simplicity, the UE height is fixed, and its location is represented by two-dimensional planar coordinates $[x, y]$. Based on the geometry-based channel model in \eqref{eq:channel}, the channel parameters $\gamma_l^{(k,s)}$, $\phi_l$, $\theta_l$ for $l \in \{1, \ldots, L\}$, as well as the number of propagation paths $L$, are conditionally dependent on the UE's location. This dependency implies that the statistical distribution of the wireless channel is inherently location-dependent.

To simplify the discussion, we consider a single-antenna UE ($\nt = 1$) and focus on the angular-delay domain representation of the channel, denoted as $\mathbf{H}_{\rm AD} \in \mathbb{C}^{\nr \times \tau}$, where $\tau < K$ represents the number of non-zero delay components retained in the delay domain \cite{li2023multi}. This representation captures key characteristics such as multipath delay and angular dispersion and is obtained by applying a two-dimensional fast Fourier transform (2D-FFT) to the spatial-frequency domain channel matrix $\mathbf{H}_{\rm SF} = [\mathbf{h}_{\rm cplx}^{(1)}, \ldots, \mathbf{h}_{\rm cplx}^{(K)}]$, where each $\mathbf{h}_{\rm cplx}^{(k)} \in \mathbb{C}^{\nr \times 1}$ denotes the channel vector corresponding to the $k$-th subcarrier.

\figref{fig:scenario} presents a comparison of angular-delay domain CSI  samples between business and rural areas. The CSI samples are simulated using the QuaDRiGa toolbox \cite{jaeckel2014quadriga} and visualized in grayscale. It can be observed that rural-area CSI exhibits sparsity due to the presence of fewer multipath clusters, while the CSI from the business area demonstrates richer features with larger angular and delay spreads, resulting from the dense scattering environment.

% \CheckRmv{
%   \begin{figure}[t]
%     \centering
%     \includegraphics[width=3.25in]{multi_region.eps}
%     \caption{angular-delay domain channel samples from subregions R1 to R5.}
%     \label{fig:multi_region}
%   \end{figure}
% }

% example: 若不放图，则表放后面
% CSI samples from different subregions are displayed in \figref{fig:multi_region}.

Next, we analyze the impact of UE speed $v$, which primarily influences the temporal variation of the wireless channel. To clearly demonstrate this effect, \figref{fig:scenario} presents a comparison of channel samples corresponding to low and high UE mobility, specifically at speeds of $v = 24$ km/h and $v = 300$ km/h, respectively. The associated frequency-time domain channels, denoted by $\mathbf{H}_{{\rm cplx}, m,n}$, are visualized as grayscale CSI maps. As observed, the CSI map corresponding to low mobility exhibits stable features along the time axis, whereas the high-mobility case results in significant temporal variations, reflecting the dynamic nature of the underlying channel conditions.
% directly impact $\gamma_l$ along the time dimension. Similarly affects the propagation and reception.

In summary, the conditioning information that characterizes the channel can be represented by the label vector $\mathbf{c} = [x, y, v]$, which encodes the UE's location and speed. Let $\mathbf{H} \in \mathbb{R}^{\nr \times \tau \times S \times 2}$ denote the real-valued channel tensor obtained by applying a 2D-FFT to the complex-valued channel $\mathbf{H}_{\rm cplx}$, followed by a concatenation of its real and imaginary components along the last dimension. For notational convenience in the proposed DM-based framework (to be detailed in \secref{sec:approach}), the tensor $\mathbf{H}$ is further vectorized into a one-dimensional representation denoted by $\mathbf{h}$.
The relationship between the channel representation $\mathbf{h}$ and the conditioning label $\mathbf{c}$ gives rise to a conditional channel distribution denoted by $q(\mathbf{h}|\mathbf{c})$. To address the issue of limited availability of measured channel data, as discussed in the previous subsection, and to support downstream tasks such as receiver processing, this work aims to learn an approximate conditional distribution $\hat{q}(\mathbf{h}|\mathbf{c})$ that closely matches the true distribution. To achieve this, a conditional DM is developed, as described in \secref{sec:approach}.
Once trained, the conditional DM can be used to generate synthetic channel samples according to $\hat{\mathbf{h}}_{\rm gen} \sim \hat{q}(\mathbf{h}|\mathbf{c})$,
% \CheckRmv{
%   \begin{equation}
%      \mathbf{h}_{\rm gen} \sim \hat{q}(\mathbf{h}|\mathbf{c}), 
%   \end{equation}
% } 
which enables the construction of augmented channel datasets tailored to specific environmental conditions defined by the label $\mathbf{c}$.
% where $\mathbf{c}$ encodes scenario-specific conditioning information. 

%%%%%%%%%%%%%%%%%%%%%%%%%%%%%%%%%%%%%%%%%%%%%%%%%%%%%%%%%%
\section{Proposed Approach} % 4 page
\label{sec:approach}

%This section presents the overall framework of the proposed method. We first describe the scenario-specific channel generation using conditional DMs, followed by the introduction of a consistency distillation strategy designed to accelerate the generation process. Finally, we discuss how the generated channels are used to augment neural receiver training. 

\subsection{Main Framework} % 0.5 page

The overall framework of the proposed approach includes the following steps: 
\begin{enumerate}
  \item \textbf{Channel Measurements:} Collect a set of $N_{\rm train}$ training samples, consisting of UE location and speed labels $\mathcal{C}_{\rm train}=\{\mathbf{c}_{\rm train}^{[j]}\}_{j=1}^{N_{\rm train}}$ and their corresponding channels $\mathcal{H}_{\rm train} = \{\mathbf{h}_{\rm train}^{[j]}\}_{j=1}^{N_{\rm train}}$. The UE's location-speed labels can be conveyed to the BS via a control link.
  Channel samples are either obtained in measurement campaigns or produced using a channel simulator. In this paper, we consider the latter and adopt the QuaDRiGa toolbox \cite{jaeckel2014quadriga} for channel simulation. To reflect practical limitations, only a limited number of samples are assumed to be available for training.
  \item \textbf{DM Training:} The DM is trained using $\mathcal{H}_{\rm train}$ by applying a forward diffusion process and learning to reverse it (denoising) while adopting $\mathcal{C}_{\rm train}$ as the conditional input. % to guide the model in generating accurate and scenario-specific channel realizations.
  The model architecture and training details are provided in \secref{sec:dm}.
  \item \textbf{Scenario-Specific Channel Generation:} Once trained, the DM is used to generate the channel dataset $\mathcal{H}_{\rm gen} =\{\hat{\mathbf{h}}_{\rm gen}^{[j]}\}_{j=1}^{N_{\rm gen}}$ given new sets of location-speed labels $\mathcal{C}_{\rm gen}=\{\mathbf{c}_{\rm gen}^{[j]}\}_{j=1}^{N_{\rm gen}}$ tailored to the scenarios intended for enhancement.
  \item \textbf{Neural Receiver Training with Augmented Dataset:} % DL models for the downstream tasks
  The final stage involves constructing a training dataset for the neural receiver using the augmented channel dataset $\mathcal{H}_{\rm train} \bigcup \mathcal{H}_{\rm gen}$. The detailed construction is presented in \secref{sec:da_sip}.
\end{enumerate}

% -- 是否全改为连续形式——能得到离散形式也有价值
\subsection{Scenario-Specific Channel Generation} \label{sec:dm}  % Conditional Generation 2 page
\subsubsection{DM-Based Conditional Generation} \label{sec:dm1} % Basics of Diffusion Models, 0.75 page
% -- principles of DM (zhou/RIS/DM生成原理讲的极简的，如channel coding)
DMs are one of the most advanced generative models, which construct a Markovian process that incrementally adds noise to diffuse structured data $\mathbf{h}\sim q_{\rm data}(\cdot)$ and learn a reverse process that generates data from noise by denoising \cite{ho2020denoising,song2020score}.  
We begin with the basic notations in these processes.
The noisy latent variables within the forward diffusion process are represented as $\{\tilde{\mathbf{h}}_t\}_{t\in [0,T]}$, where $t$ denotes the time variable, and $T$ is a fixed constant representing the time range. 
Let $q_t(\cdot)$ denote the distribution of $\tilde{\mathbf{h}}_t$. The initial distribution of the forward diffusion process is given by $q_{0}(\cdot) = q_{\rm data}(\cdot)$, and the final distribution $q_T(\cdot)$ is an isotropic Gaussian.  % distribution of $\mathbf{h}_0$ and $\mathbf{h}_T$
The noise added in the forward diffusion process at time $t$ is denoted as $\boldsymbol{\epsilon}_t$, which is approximated using a noise prediction NN $\boldsymbol{\epsilon}_{\boldsymbol{\theta}}(\tilde{\mathbf{h}}_t, t)$ parameterized by $\boldsymbol{\theta}$ in the reverse process for denoising and generation.
% The architecture of this network can be flexibly designed and is specifically crafted in the following.

The diffusion process can be represented as a stochastic differential equation (SDE) given by   
\CheckRmv{
  \begin{equation}
    \mathrm{d}\tilde{\mathbf{h}}_t = \boldsymbol{\mu}(\tilde{\mathbf{h}}_t, t)\mathrm{d}t+g(t) \mathrm{d}\mathbf{w}_t,
  \end{equation}
}
where $\boldsymbol{\mu}(\cdot, \cdot)$ and $g(\cdot)$ denote the drift and diffusion coefficients, respectively, and $\mathbf{w}_t$ is a standard Brownian process. A distinct feature of this SDE is that it possesses a probability-flow ordinary differential equation (PF-ODE) as the reverse process, whose solution trajectory distribution at time $t$ aligns with the marginal distribution $q_t(\cdot)$:
\CheckRmv{
  \begin{equation}
    \mathrm{d}\hat{\mathbf{h}}_t = \left[\boldsymbol{\mu}(\hat{\mathbf{h}}_t, t) - \frac{1}{2}g^2(t) \nabla_{\hat{\mathbf{h}}} \log q_t(\hat{\mathbf{h}}_t) \right] \mathrm{d}t,
  \end{equation}
}
where {$\{\hat{\mathbf{h}}_t\}_{t\in [0,T]}$ denotes the denoised channel variables within the reverse diffusion process}, and $\nabla_{\hat{\mathbf{h}}} \log q_t(\hat{\mathbf{h}}_t)$ represents the score function of $q_t(\hat{\mathbf{h}}_t)$, pointing to the high-density region of data. 

In alignment with established practices presented in \cite{karras2022elucidating}, we set $\boldsymbol{\mu}(\hat{\mathbf{h}}_t,t) = \mathbf{0}$ and $g(t)=\sqrt{2t}$ in this work. Under this configuration, the noise schedule in the diffusion process becomes $\sigma(t) = t$, leading to the conditional distribution of the noisy latent variable $\tilde{\mathbf{h}}_t$ being expressed as $q(\tilde{\mathbf{h}}_t|\mathbf{h})=\mathcal{N}(\tilde{\mathbf{h}}_t;\mathbf{h}, t^2\mathbf{I})$ \cite{song2020score,karras2022elucidating,pei2025latent}. 
To numerically solve the PF-ODE, the time range $t\in [0, T]$ is split into a series of time intervals with $(N+1)$ boundaries, denoted as $\epsilon=t_0<\ldots<t_N=T$. Herein, $\epsilon$ is a predefined small positive constant introduced to avoid numerical instability. Specifically, we follow \cite{karras2022elucidating} to set $t_0=\epsilon=0.002$, $t_N=T=80$, and 
\CheckRmv{
  \begin{equation}
    t_n =\left(t_0^{1/\omega} + \frac{n}{N}(t_n^{1/\omega} - t_0^{1/\omega})\right)^{\omega} 
  \end{equation}
} 
with $\omega=7$.

% {\rl The conditional distribution of the latent variables $\mathbf{h}_t$ given $\mathbf{h}_0$ can be written as
% \CheckRmv{
%   \begin{equation}
%     q(\mathbf{h}_t | \mathbf{h}_0) = \mathcal{N}(\mathbf{h}_t;\mathbf{h}_0, t^2\mathbf{I}),
%   \end{equation}
% }}
To enable the reverse process for sample generation, we should further have an approximation of the score function $\nabla_{\hat{\mathbf{h}}} \log q_t(\hat{\mathbf{h}}_t)$. This approximation can be achieved by denoising score matching \cite{vincent2011connection}, which leverages a denoiser $\mathbf{d}_{\boldsymbol{\theta}}(\mathbf{h}_t, t)$ to estimate the score:
\CheckRmv{
  \begin{equation}
    \nabla_{\hat{\mathbf{h}}} \log q_t(\hat{\mathbf{h}}_t) \approx \frac{\mathbf{d}_{\boldsymbol{\theta}}(\hat{\mathbf{h}}_t, t) - \hat{\mathbf{h}}_t}{t^2}. 
    \label{eq:dsm}
  \end{equation}
} 
The denoiser is optimized by minimizing the expectation of the $l_2$ denoising error $\|\mathbf{d}_{\boldsymbol{\theta}}(\hat{\mathbf{h}}_t, t) - \mathbf{h}\|_2^2$ for samples $\mathbf{h}$ drawn from $q_{\rm data}(\cdot)$.

% an NN is trained to approximate the score function, denoted as $s_{\boldsymbol{\theta}}(\mathbf{h}_t,t)\approx \nabla_{\mathbf{h}} \log q_t(\mathbf{h}_t)$. This score model is typically implemented by the noise prediction model $\boldsymbol{\epsilon}_{\boldsymbol{\theta}}(\mathbf{h}_t,t)$ adopted in the DM's denoiser as: $s_{\boldsymbol{\theta}}(\mathbf{h}_t, t) = - \boldsymbol{\epsilon}_{\boldsymbol{\theta}}(\mathbf{h}_t,t) / t$ \cite{song2020score,luo2022understanding}. 
Given the above configurations and approximation, an empirical estimate of the PF-ODE for sampling can be expressed as
\CheckRmv{
  \begin{equation}
    \frac{{\rm d} \hat{\mathbf{h}}_t}{{\rm d}t} = \frac{\hat{\mathbf{h}}_t - \mathbf{d}_{\boldsymbol{\theta}}(\hat{\mathbf{h}}_t, t)}{t}. 
  \end{equation}
}
% % generation总体描述  % 可能不给具体公式，CM中再说
% {To solve this empirical PF-ODE, the Heun solver is utilized \cite{karras2022elucidating}, denoted as 
% \CheckRmv{
%   \begin{align}
%     \left. \frac{{\rm d} \mathbf{h}_t}{{\rm d}t}\right|_{t=t_{n+1}} &\approx \frac{\mathbf{h}_{t_{n+1}}-\mathbf{h}_{t_{n}}}{t_{n+1}-t_{n}} \nonumber \\
%      &\approx \frac{1}{2}\left(\boldsymbol{\epsilon}_{\boldsymbol{\theta}}(\mathbf{h}_{t_n}, t_n) + \boldsymbol{\epsilon}_{\boldsymbol{\theta}}(\mathbf{h}_{t_{n+1}}, t_{n+1})\right). 
%   \end{align}
% }
% Therefore, the estimate of $\mathbf{h}_{t_n}$ is given by
% \CheckRmv{
%   \begin{align}
%     \hat{\mathbf{h}}_{t_n}^{\boldsymbol{\theta}} \approx \mathbf{h}_{t_{n+1}}- \frac{1}{2} &\left(\boldsymbol{\epsilon}_{\boldsymbol{\theta}}(\mathbf{h}_{t_n}, t_n) + \boldsymbol{\epsilon}_{\boldsymbol{\theta}}(\mathbf{h}_{t_{n+1}}, t_{n+1}) \right) \nonumber  \\ 
%     &(t_{n+1} - t_n), 
%   \end{align}
% }
% where $\mathbf{h}_{t_n}$ can be further approximated by $\mathbf{h}_{t_{n+1}} - \boldsymbol{\epsilon}_{\boldsymbol{\theta}}(\mathbf{h}_{t_{n+1}}, t_{n+1})(t_{n+1}-t_n)$ based on the Euler approximation \cite{song2020score}.}
% 
This empirical PF-ODE can be solved using the Euler \cite{song2020score} or Heun \cite{karras2022elucidating} numerical ODE solvers, providing the sampling trajectory from $\hat{\mathbf{h}}_T \sim \mathcal{N}(\mathbf{0},T^2\mathbf{I})$ to $\hat{\mathbf{h}}_{\epsilon}$. %\endnote{{\rl With a little abuse of notations, we use the 2D Gaussian distribution to indicate}}  
The final estimate $\hat{\mathbf{h}}_{\epsilon}$ converges to an approximate sample from the target data distribution $q_{\rm data}(\cdot)$.
%generation---实际采样用的[Karras, Alg.~2].

% -- conditional & scenario-specific generation (RF-diffusion/Gong/Andrews)
In our proposed approach, a DM is employed to generate channel tensors $\mathbf{h}$ conditioned on the UE location-speed vector $\mathbf{c}$, thereby capturing environment-specific characteristics through conditional generation. 
Specifically, $\mathbf{c}$ serves as the conditional input that guides the reverse diffusion process, represented as $q_{\boldsymbol{\theta}}(\hat{\mathbf{h}}_{t_{n-1}}|\hat{\mathbf{h}}_{t_n},\mathbf{c})$.
To incorporate this conditional dependency, the denoiser $\mathbf{d}_{\boldsymbol{\theta}}$ is designed to accept $\mathbf{c}$ as a conditional input, denoted as $\mathbf{d}_{\boldsymbol{\theta}}(\cdot,\cdot;\mathbf{c})$. Furthermore, to enhance training stability, we construct the denoiser using a preconditioned architecture that integrates a skip connection and the noise prediction model $\boldsymbol{\epsilon}_{\boldsymbol{\theta}}$ as
\CheckRmv{
  \begin{equation}
    \mathbf{d}_{\boldsymbol{\theta}}(\hat{\mathbf{h}}_t,t;\mathbf{c}) = c_{\rm skip}(t)\hat{\mathbf{h}}_t + c_{\rm out}(t)\boldsymbol{\epsilon}_{\boldsymbol{\theta}}(\hat{\mathbf{h}}_t,t;\mathbf{c}), \label{eq:nn_precondition}
  \end{equation}
}
where $c_{\rm skip}(t)$ is used to modulate the skip connection, and $c_{\rm out}(t)$ scales the noise prediction model's output. {The selection of these two functions is given by
\CheckRmv{
  \begin{equation}
    c_{\rm skip}(t)= \frac{\sigma_{\rm d}^2}{\sigma_{\rm d}^2 + (t-\epsilon)^2}, \; c_{\rm out}(t) = \frac{(t-\epsilon)\cdot \sigma_{\rm d}}{\sqrt{\sigma_{\rm d}^2 + t^2}}, \label{eq:cskip_cout}
  \end{equation}
}
where $\sigma_{\rm d}$ is empirically set to 0.5 based on a parameter scan over the range $[0.1, 1.0]$. This value achieves a trade-off between gradient stability and convergence speed, thereby balancing the two weighting coefficients and ensuring stable training.} This configuration ensures that the output of $\boldsymbol{\epsilon}_{\boldsymbol{\theta}}$ maintains unit variance, which effectively stabilizes gradient magnitudes during training regardless of varying noise levels.
Consequently, the final distribution of the constructed reverse process provides an accurate approximation of the true conditional channel distribution $q(\mathbf{h}|\mathbf{c})$, facilitating scenario-aware channel sample generation.

% -- loss function and training   %可能按denoising score matching来讲
% The objective of learning the reverse process is to train the parameters $\boldsymbol{\theta}$, % which can also be 
% making the predicted noise % of size $\mathbb{R}^{\nr \times \tau \times S \times 2}$
% from $\boldsymbol{\epsilon}_{\boldsymbol{\theta}}$ closely approaches the true noise $\boldsymbol{\epsilon}_{t_n}$ incorporated at each time boundary $t_n$ by the numerical solver. Hence, the loss function can be written as \cite{ho2020denoising}. 
Given the denoising score matching approximation in \eqref{eq:dsm}, the objective function of learning the reverse process can be written as \cite{vincent2011connection} 
\CheckRmv{
  \begin{equation}
    \begin{aligned}
      % \mathbb{E}_{\mathbf{h}_0, \boldsymbol{\epsilon}_{t_n}, n}\left[\|\boldsymbol{\epsilon}_{t_n} - \boldsymbol{\epsilon}_{\boldsymbol{\theta}}(\mathbf{h}_{t_n},t_n;\mathbf{c})\|_2^2 \right], \\
      \mathcal{L}_{\boldsymbol{\theta}}=\mathbb{E}_{\mathbf{h},\hat{\mathbf{h}}_{t_n},n}\left[\|\mathbf{d}_{\boldsymbol{\theta}}(\hat{\mathbf{h}}_{t_n}, t_n;\mathbf{c}) - {\mathbf{h}}\|_2^2\right],
    \end{aligned}
  \end{equation}
}
where $\mathbf{h}\sim q_{\rm data}(\cdot)$, $\hat{\mathbf{h}}_{t_n}\sim \mathcal{N}(\mathbf{h},t_n^2\mathbf{I})$, and $n\sim \mathcal{U}(\{1,\ldots,N\})$. 
% We also apply a time-dependent weighting to this objective according to \cite{karras2022elucidating}.
Considering the relationship between the denoiser and the noise prediction model $\boldsymbol{\epsilon}_{\boldsymbol{\theta}}$, this objective can be equivalently reformulated as minimizing the error between the predicted noise $\boldsymbol{\epsilon}_{\boldsymbol{\theta}}(\hat{\mathbf{h}}_{t_n},t_n;\mathbf{c})$ and the true noise $\boldsymbol{\epsilon}_{t_n}$ injected by the forward SDE at any time $t_n$ ($n=1,\ldots,N$). 
% The input of the NN contains three components: the noisy latent variable $\mathbf{h}_t$, the time step index $t$, and the conditioning $\mathbf{c}$, enabling progressively accurate characterization of the channel.

\CheckRmv{
  \begin{figure}[t]
    \centering
    \includegraphics[width=3in]{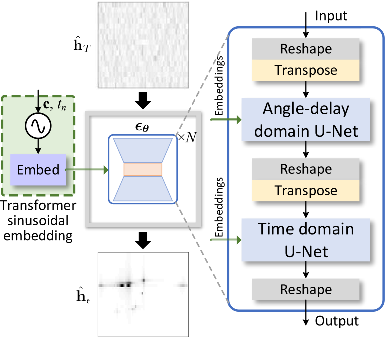}
    \caption{Conditional generation process using the NN $\boldsymbol{\epsilon}_{\boldsymbol{\theta}}$. The channel samples $\hat{\mathbf{h}}_T$ and $\hat{\mathbf{h}}_{\epsilon}$ are illustrated as 2D grayscale images for clarity.}
    \label{fig:inference}
  \end{figure}
}

% 放到conditional generation小节
\renewcommand{\algorithmicrequire}{\textbf{Input:}}
\renewcommand{\algorithmicensure}{\textbf{Output:}}
\renewcommand{\algorithmiccomment}[1]{/* #1 */}
\newcommand{\IfThen}[2]{% \IfThenElse{<if>}{<then>}
  \STATE \algorithmicif\ #1\ \algorithmicthen\ #2}
\newcommand{\parfor}[1]{\STATE \algorithmicfor\ #1  \textbf{do in parallel}}
% \algsetup{indent=0.5em}
\CheckRmv{
  \begin{algorithm}[t]
    \caption{DM-Based Conditional Generation} 
    \label{alg:sampling}
    % \small
    \begin{algorithmic}[1] 
      \REQUIRE $\boldsymbol{\epsilon}_{\boldsymbol{\theta}}(\cdot,\cdot;\cdot)$, $c_{\rm skip}(t)$, $c_{\rm out}(t)$, $t_{n\in \{0,\cdots,N\}}$, $\mathcal{C}_{\rm gen}$. 
      \STATE \textbf{Initialize:} $\hat{\mathbf{h}}_{t_N}=\hat{\mathbf{h}}_{T}\sim \mathcal{N}(\mathbf{0},T^2\mathbf{I})$, $\mathbf{c}\sim\mathcal{C}_{\rm gen}$.
      \FOR{$n=N$ to $1$}
      \STATE Compute $\mathbf{d}_{\boldsymbol{\theta}}(\hat{\mathbf{h}}_{t_n}, t_n;\mathbf{c})$ using \eqref{eq:nn_precondition}.
      \STATE $\boldsymbol{d}_n \gets \frac{\hat{\mathbf{h}}_{t_n} - \mathbf{d}_{\boldsymbol{\theta}}(\hat{\mathbf{h}}_{t_n}, t_n;\mathbf{c})}{t_n}$.
      \STATE $\hat{\mathbf{h}}^{\prime}_{t_{n-1}} \gets \hat{\mathbf{h}}_{t_n} + (t_{n-1} - t_n) \boldsymbol{d}_n$.
      \STATE Compute $\mathbf{d}_{\boldsymbol{\theta}}(\hat{\mathbf{h}}^{\prime}_{t_{n-1}}, t_{n-1};\mathbf{c})$ using \eqref{eq:nn_precondition}.
      \STATE $\boldsymbol{d}_n^{\prime} \gets \frac{\hat{\mathbf{h}}^{\prime}_{t_{n-1}} - \mathbf{d}_{\boldsymbol{\theta}}(\hat{\mathbf{h}}^{\prime}_{t_{n-1}}, t_{n-1};\mathbf{c})}{t_{n-1}}$.
      \STATE $\hat{\mathbf{h}}_{t_{n-1}} \gets \hat{\mathbf{h}}_{t_n} + (t_{n-1} - t_n)(\frac{1}{2}\boldsymbol{d}_{n} + \frac{1}{2}\boldsymbol{d}_{n}^{\prime})$.
      \ENDFOR
      \ENSURE $\hat{\mathbf{h}}_{\epsilon}=\hat{\mathbf{h}}_{t_0}$. 
    \end{algorithmic} 
  \end{algorithm}
}

% -- conditional generation process 
The conditional generation process is presented in \figref{fig:inference}, which produces a synthetic channel sample $\hat{\mathbf{h}}_{\epsilon}$ from the pure Gaussian noise $\hat{\mathbf{h}}_{T}$ given $\mathbf{c}$ and $t$ through $N$ iterations of the ODE solver with $t=t_1,\ldots,t_N$. 
% todo: 实际采样用的[Karras, Alg.~2]，有stochasticity，只一句话带过？
At iteration $n$, the network outputs the estimated noise $\boldsymbol{\epsilon}_{\boldsymbol{\theta}}(\hat{\mathbf{h}}_{t_n}, t_n;\mathbf{c})$, % three inputs
which is utilized to construct $\hat{\mathbf{h}}_{t_{n-1}}$ based on the ODE solver.
We adopt the Heun solver in our implementation. Details of the sampling algorithm are summarized in \algref{alg:sampling} and clarified in the Appendix.  %{/ as elaborated in \cite{karras2022elucidating}.} %{To further enhance the generation quality, stochasticity is introduced %by strategically adding noise  
%at each iteration, following the approach in \cite{karras2022elucidating}.} 
% {In practice, it is observed that introducing stochasticity in the solver enhances the overall generation quality \cite{karras2022elucidating}. Reverse generative SDE.
% The stochastic solver adds and removes noise at each iteration. 
% Algorithm for the adopted Heun solver.}
Each generated $\hat{\mathbf{h}}_{\epsilon}$ is collected as $\hat{\mathbf{h}}_{\rm gen}^{[j]}$ to constitute $\mathcal{H}_{\rm gen}$ for the subsequent augmentation.

% dual U-Nets
\figref{fig:inference} also illustrates the architectural design of $\boldsymbol{\epsilon}_{\boldsymbol{\theta}}$. 
To effectively generate 3D CSI samples that align with the true channel statistics, the NN architecture is customized beyond conventional designs that utilize a standard U-Net structure \cite{ronneberger2015unet}. 
The proposed architecture consists of two cascaded subnetworks, denoted as $\boldsymbol{\epsilon}^{\prime}_{\boldsymbol{\theta}_1}$ and $\boldsymbol{\epsilon}^{\prime}_{\boldsymbol{\theta}_2}$, each adopting the U-Net structure as its backbone.
Since a standard U-Net operates on two-dimensional data, the input tensor $\hat{\mathbf{h}}_t$ is reshaped and permuted into two formats: $BS \times \nr \times \tau \times 2$ and $B \times \nr\tau \times S \times 2$, where $B$ denotes the batch size. These two representations are then fed into $\boldsymbol{\epsilon}^{\prime}_{\boldsymbol{\theta}1}$ and $\boldsymbol{\epsilon}^{\prime}_{\boldsymbol{\theta}_2}$, respectively.

The first subnetwork focuses on extracting spatial features in the angular-delay domain, while the second captures temporal variations across consecutive symbols. Although both subnetworks share the same architectural configuration, they are trained with independent parameter sets, $\boldsymbol{\theta}_1$ and $\boldsymbol{\theta}_2$, to specialize in their respective domains. This architecture extends the applicability of diffusion models from conventional two-dimensional data to the generation of three-dimensional channel tensors by jointly modeling spatial and temporal characteristics. To enable conditional generation, the conditioning label $\mathbf{c}$, along with the diffusion time step $t_n$, is incorporated into the network via Transformer-style sinusoidal positional encoding \cite{vaswani2017attention}, followed by a dense embedding layer.

\CheckRmv{
  \begin{figure}[t]
    \centering
    \includegraphics[width=3.25in]{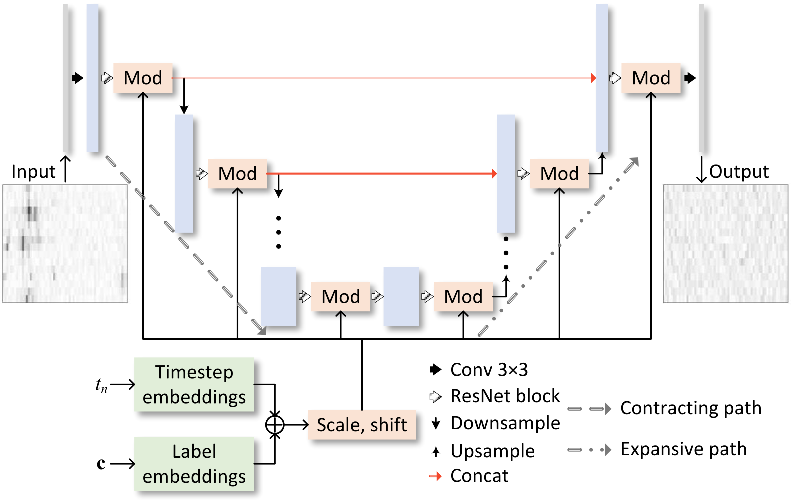}
    \caption{U-Net architecture of the subnetwork.}
    \label{fig:network}
  \end{figure}
}

% -- network structure: angular-delay and time-domain U-Net
% Regarding the architecture of the subnetworks, we adopt the U-Net structure \cite{ronneberger2015unet} as the backbone, which contains multiple convolutional layers at different resolutions. % a convolutional NN (CNN) with encoder-decoder structure. 
% % -- U-Net简介  参照RIS（篇幅）、Gong、网络结构、U-Net描述文章
% The detailed structure of this network is illustrated in \figref{fig:network}, 
The architecture of the U-Net used in the subnetworks is illustrated in \figref{fig:network}. It consists of a contracting path and an expansive path across multiple resolution levels. % resblock数和downsampling/upsampling layer数的关系
% contracting path
The contracting path is designed to extract multi-scale features from the input through a sequence of residual blocks based on the ResNet architecture \cite{he2016deep}. Each block includes convolutional layers, activation functions, and residual connections. To enhance spatial feature learning, self-attention modules are inserted at selected downsampling stages.
The expansive path reconstructs the noise prediction output from the intermediate features using a sequence of residual blocks, each followed by an upsampling layer that doubles the resolution of the feature map. This structure enables the recovery of fine-grained spatial information from compressed feature representations.
To enable conditional generation, the timestep and scenario label $\mathbf{c}$ are embedded and transformed into scale and shift vectors. These vectors modulate the activations within the ResNet blocks, allowing the network to adapt to different channel conditions.

\subsubsection{Consistency Distillation for Acceleration}  % for acceleration 1 page
%介绍问题（和现有方法）
A major limitation of DM-based generation is the slow inference speed, as the process typically requires $N$ iterative steps, often in the order of hundreds. This subsection introduces a scheme that significantly reduces the number of required steps while preserving generation quality. %

To accelerate generation, we utilize the consistency distillation strategy \cite{song2023consistency} to achieve one-step generation without significantly compromising the quality. This strategy aims at constructing a consistency model (CM) that can map any latent variables $\hat{\mathbf{h}}_t$ on the same trajectory to a consistent origin $\hat{\mathbf{h}}_{\epsilon}$.
Specifically, this property can be represented by a consistency function $\mathbf{f}_{{\boldsymbol{\theta}}}$ satisfying
\CheckRmv{
  \begin{equation}
    \mathbf{f}_{{\boldsymbol{\theta}}}(\hat{\mathbf{h}}_t, t;\mathbf{c}) = \mathbf{f}_{{\boldsymbol{\theta}}}(\hat{\mathbf{h}}_{t^{\prime}}, t^{\prime}; \mathbf{c}),\; \forall t, t^{\prime} \in [\epsilon, T]. \label{eq:consistency_func}
  \end{equation}
}
With this property, one-step generation to synthesize the target data can be achieved through $\hat{\mathbf{h}}_{\epsilon}=\mathbf{f}_{\boldsymbol{\theta}}(\hat{\mathbf{h}}_T, T;\mathbf{c})$.

We derive the CM by distilling the denoiser $\mathbf{d}_{\boldsymbol{\theta}}$ in the teacher DM trained in the previous subsection.
A notable reason for choosing $\mathbf{d}_{\boldsymbol{\theta}}$ as the teacher is that its parameterization in \eqref{eq:nn_precondition} and \eqref{eq:cskip_cout} naturally satisfies the boundary condition $\mathbf{f}_{\boldsymbol{\theta}}(\hat{\mathbf{h}}_{\epsilon}, \epsilon;\mathbf{c})=\hat{\mathbf{h}}_{\epsilon}$ implied by \eqref{eq:consistency_func} since $c_{\rm skip}(\epsilon)=1$ and $c_{\rm out}(\epsilon)=0$.
The distillation can be achieved by minimizing the following objective 
\CheckRmv{
  \begin{align}
      \mathcal{L}&({\boldsymbol{\theta}},\boldsymbol{\theta}^{-})  \nonumber \\ 
      = \mathbb{E}&[\|\mathbf{f}_{\boldsymbol{\theta}}(\hat{\mathbf{h}}_{t_{n+1}}, t_{n+1};\mathbf{c}) - \mathbf{f}_{{\boldsymbol{\theta}}^{-}}(\hat{\mathbf{h}}_{t_n}^{\Phi}, t_n;\mathbf{c})\|_2^2],
      \label{eq:cm_loss}
  \end{align}
}
where the expectation is taken over $\mathbf{h}\sim q_{\rm data}(\cdot)$, $n\sim \mathcal{U}(\{0,\ldots,N-1\})$, and $\hat{\mathbf{h}}_{t_{n+1}}\sim \mathcal{N}(\mathbf{h},t_{n+1}^2\mathbf{I})$. ${\boldsymbol{\theta}}^{-}$ is a running average of the preceding values of $\boldsymbol{\theta}$ during the course of training \cite{song2023consistency}. $\hat{\mathbf{h}}_{t_n}^{\Phi}$ is an estimate of $\hat{\mathbf{h}}_{t_n}$ from $\hat{\mathbf{h}}_{t_{n+1}}$ by adopting one step of the numerical ODE solver $\Phi$, %(e.g.~the Euler or Heun solvers, whose update details can be found in \cite{karras2022elucidating,pei2025latent}), 
given by
\CheckRmv{
  \begin{equation}
    \hat{\mathbf{h}}_{t_n}^{\Phi} = \hat{\mathbf{h}}_{t_{n+1}} + (t_n - t_{n+1})\Phi(\hat{\mathbf{h}}_{t_{n+1}}, t_{n+1}). 
    \label{eq:onestep}
  \end{equation}
} 
Similar to the DM-based sampling elucidated in \secref{sec:dm1}, the Heun solver is utilized, and the update rules of the corresponding $\Phi$ can be found in \algref{alg:sampling}.  %is a function of the pre-trained $\boldsymbol{\epsilon}_{\boldsymbol{\theta}}$. The update details are presented in the Appendix.

The parameters $\boldsymbol{\theta}$ are updated based on gradient descent over \eqref{eq:cm_loss} with a learning rate $\eta=10^{-5}$, while $\boldsymbol{\theta}^{-}$ is refined after each gradient descent step using exponential moving average: 
\CheckRmv{
  \begin{equation}
    {\boldsymbol{\theta}}^{-} = \mathrm{stopgrad}(\beta \boldsymbol{\theta}^{-} + (1-\beta)\boldsymbol{\theta}),
  \end{equation}
}
where the operation $\mathrm{stopgrad}(\cdot)$ blocks the gradient propagation to enhance the stability in training, and $\beta$ denotes the decay rate, which is empirically chosen as $0.95$ in this work.

The training procedure of consistency distillation is outlined in \algref{alg:cd}.
As $\boldsymbol{\theta}^{-}$ represents an average of $\boldsymbol{\theta}$ during training, these parameters converge to identical values ($\boldsymbol{\theta}^{-} = \boldsymbol{\theta}$) upon reaching equilibrium.
Once trained, the distilled CM can produce target samples $\hat{\mathbf{h}}_{\epsilon} $ from Gaussian-distributed inputs $\hat{\mathbf{h}}_T$ at time $T$ through the deterministic mapping $\hat{\mathbf{h}}_{\epsilon} = \mathbf{f}_{\boldsymbol{\theta}}(\hat{\mathbf{h}}_T, T;\mathbf{c})$.

% \algsetup{indent=0.5em}
\CheckRmv{
  \begin{algorithm}[t]
    \caption{Consistency Distillation}  
    \label{alg:cd}
    % \small
    \begin{algorithmic}[1] 
      \REQUIRE 
      $\mathcal{H}_{\rm train}$, $\mathcal{C}_{\rm train}$, $\eta$, $\beta$, $\Phi(\cdot,\cdot)$, initial model parameter $\boldsymbol{\theta}$.
      \STATE \textbf{Initialize:} $\boldsymbol{\theta}^{-} \gets \boldsymbol{\theta}$; $\hat{\mathbf{h}}_{T}\sim \mathcal{N}(\mathbf{0},T^2\mathbf{I})$.
      \REPEAT
      \STATE Sample $\mathbf{h}\sim \mathcal{H}_{\rm train}$, $\mathbf{c}\sim\mathcal{C}_{\rm train}$ and $n\sim \mathcal{U}(\{0,\ldots, N-1\})$.
      \STATE Sample $\hat{\mathbf{h}}_{t_{n+1}}\sim\mathcal{N}(\mathbf{h},t_{n+1}^2\mathbf{I})$.
      \STATE Compute $\hat{\mathbf{h}}_{t_n}^{\Phi}$ using \eqref{eq:onestep}.
      \STATE Compute $\mathcal{L}({\boldsymbol{\theta}},\boldsymbol{\theta}^{-})\gets \|\mathbf{f}_{\boldsymbol{\theta}}(\hat{\mathbf{h}}_{t_{n+1}}, t_{n+1};\mathbf{c}) - \mathbf{f}_{{\boldsymbol{\theta}}^{-}}(\hat{\mathbf{h}}_{t_n}^{\Phi}, t_n;\mathbf{c})\|_2^2$.
      \STATE Compute $\boldsymbol{\theta} \gets \boldsymbol{\theta} - \eta \nabla_{\boldsymbol{\theta}}\mathcal{L}({\boldsymbol{\theta}},\boldsymbol{\theta}^{-})$.
      \STATE Compute $\boldsymbol{\theta}^{-} \gets \mathrm{stopgrad}(\beta \boldsymbol{\theta}^{-} + (1-\beta)\boldsymbol{\theta})$.
      \UNTIL{convergence}
      \ENSURE Distilled CM $\mathbf{f}_{\boldsymbol{\theta}}(\cdot,\cdot)$. 
    \end{algorithmic} 
  \end{algorithm}
}

\subsection{Data Augmentation for the SIP Neural Receiver} \label{sec:da_sip}  % 0.25 page 
We adopt the neural receiver proposed in \cite{xiao2025interference} for SIP transmission. The neural receiver contains two ResNet-based NNs targeted channel estimation and data detection, respectively, and adopts the interference cancellation scheme to enhance accuracy.

% The aim of training the receiver is to minimize 
The input-output relationship of the neural receiver function $R$, which involves $I$ interference cancellation iterations between the channel estimator and the data detector, is given by 
\CheckRmv{
  \begin{equation}
    \hat{\mathbf{B}}_i, \hat{\mathbf{H}}_{{\rm SF}, i} = R(\mathbf{Y}, \mathbf{P}),\; 1\leq i \leq I, 
  \end{equation}
}
where $\mathbf{Y}\in \mathbb{R}^{\nr \times K \times S \times 2}$ and $\mathbf{P}\in \mathbb{R}^{\nt \times K \times S \times 2}$ represent the real-valued received signal tensor and pilot tensor, respectively. $\hat{\mathbf{B}}_i \in \mathbb{R}^{\nt \times K \times S \times Q}$ and $\hat{\mathbf{H}}_{{\rm SF}, i} \in \mathbb{R}^{\nr \times \nt \times K \times S \times 2}$ denote the (soft) estimated bits and the estimated channel in the spatial-frequency domain, respectively, where $Q$ is the number of bits per symbol corresponding to the quadrature amplitude modulation (QAM) order.
The training objective of the neural receiver can be written as 
\CheckRmv{
  \begin{equation}
    \begin{aligned}
      \min_{\boldsymbol{\Theta}}\; & \frac{1}{I}\sum_{i=1}^{I}\left\{\mathcal{L}_{\mathrm{BCE}}({\mathbf{B}},\hat{\mathbf{B}}_{i})+\mathcal{L}_{\mathrm{MSE}}(\mathbf{H}_{\rm SF},\hat{\mathbf{H}}_{{\rm SF},i})\right\}, \\
    \end{aligned}
  \end{equation}
}
where $\boldsymbol{\Theta}$ is the parameter set of the neural receiver. $\mathcal{L}_{\rm BCE}$ and $\mathcal{L}_{\rm MSE}$ denote the binary cross entropy (BCE) and mean square error (MSE) loss. ${\mathbf{B}}$ and $\mathbf{H}_{{\rm SF}}$ are the encoded bits and the true channel in the spatial-frequency domain. 

The augmented dataset for training the neural receiver can be divided into pairs of the form $\{\mathbf{Y}^{[j]},\mathbf{P}^{[j]},\mathbf{B}^{[j]},\mathbf{H}_{\rm SF}^{[j]}\}$. To construct this dataset, the channel tensors within $\mathcal{H}_{\rm train} \bigcup \mathcal{H}_{\rm gen}$ are first transformed into the spatial-frequency domain to obtain $\mathbf{H}_{\rm SF}^{[j]}$ using a 2D-inverse FFT. Subsequently, the pilot tensor $\mathbf{P}^{[j]}$ and bit tensor $\mathbf{B}^{[j]}$ are randomly generated. The corresponding received signal tensor $\mathbf{Y}^{[j]}$ is then derived based on \eqref{eq:receive}, with SNRs randomly selected from a predefined range.

\CheckRmv{
  \begin{table}[t]
    \centering
    \begin{threeparttable}
    \caption{Parameter Settings in the Channel Simulation}
    \setlength\tabcolsep{2.8pt}
    \begin{tabular}{l|l} 
    \hline \hline
    \multirow{2}{*}{Antenna setting}  & BS: $\nr=32$ ULA antennas          \\
                                      & UE: single antenna ($\nt=1$)  \\
    \hline
    Center frequency                  & 2.655 GHz           \\
    Bandwidth                         & 10 MHz              \\
    FFT size (subcarriers)            & $K=512$                 \\ 
    OFDM symbol number                & $S=14$   \\
    Delay domain length               & $\tau=32$ \\
    \hline
    Subregion range                   & 20 m $\times$ 20 m \\
    UE speed                          & 24 km/h, 300 km/h \\
    \hline
    {Pre-augmentation channel} & 1000 (200 samples per subregion) \\
    samples                    & Evenly split between 24 km/h and 300 km/h \\
    \hline \hline
    \end{tabular}
    \label{tab:setup}
  \end{threeparttable}
  \end{table}
}

\CheckRmv{
  \begin{table}[t]
    \centering
    \begin{threeparttable}
    \caption{Descriptions of Different Subregions}
    \setlength\tabcolsep{1pt}
    \begin{tabular}{lccc} 
    \hline \hline
    Subregion            &{Center position[m]}      &Scenario        &Clusters    \\ 
    \hline
    R1 business area   &(50, 0)              &3GPP-38.901-UMi-NLOS
    &40                      \\
    R2 residential area  &(-100, -50)          &3GPP-38.901-UMi-NLOS
    &40                \\
    R3 rural area        &(10, -70)            &3GPP-38.901-UMi-LOS
    &5              \\
    R4 parking lot       &(90, -160)           &3GPP-38.901-UMi-LOS
    &5              \\
    R5 warehouse         &(0, 170)             &3GPP-38.901-UMi-NLOS
    &10              \\
    \hline \hline
    \end{tabular}
    \label{tab:region}
  \end{threeparttable}
  \end{table}
}

\CheckRmv{
      \begin{table}[t]
        \centering
        {\begin{threeparttable}
        \caption{{Hyperparameters of Each U-Net Subnetwork}}
        \setlength\tabcolsep{3pt}
        \begin{tabular}{l|cccc} 
        \hline \hline
        Hyperparameter      &Channel dim.  & Multipliers  &\#ResNet blocks & \#Params        \\ 
        \hline
        {Value}        &128                &\{1, 2, 3, 4\}     &1  & 72.43M \\
        \hline \hline
        \end{tabular}
        \label{tab:unet_para}
        \begin{tablenotes}[para,flushleft]
          \footnotesize
          Note: {Channel dim. denotes the depth of the first convolutional layer. Multipliers are applied to the depth of subsequent resolution levels.}
        \end{tablenotes}
      \end{threeparttable}}
      \end{table}
    }

%%%%%%%%%%%%%%%%%%%%%%%%%%%%%%%%%%%%%%%%%%%%%%%%%%%%%%%%%%
\section{Numerical Results}  % 2.5-3 page
\label{sec:simu}

%This section first introduces the simulation setup, followed by an evaluation and analysis of the performance of the proposed methods.

\subsection{Simulation Setups}  \label{sec:simu_setup}

We utilize the QuaDRiGa software tool \cite{jaeckel2014quadriga} to simulate the wireless channels. The simulation parameter settings are summarized in \tabref{tab:setup}.
Following the setup in \cite{li2023multi}, five representative subregions are modeled within an urban microcell (UMi) environment, as detailed in \tabref{tab:region}. Each subregion is a square area measuring 20 m by 20 m, centered at the specified coordinates. The BS is located at the origin $(0, 0)$, and UEs are uniformly distributed within the subregions and follow linear movement trajectories. 
% Similar to \cite{li2023multi}, we simulate an urban cell containing multiple subregions as shown in \tabref{tab:region}. Each UE is randomly dropped in one of these subregions. 
%
A training dataset consisting of 1000 channel samples ($N_{\rm train} = 1000$) is used to train the generative models prior to channel synthesis and augmentation. 
% 数据描述
Specifically, 200 samples are generated per subregion, with an equal division between UEs moving at 24 km/h and 300 km/h.

The U-Net architectures used in both the DM and the CM have an initial convolutional depth of 128. The depth multipliers across the four resolution levels are set to {1, 2, 3, 4}, and each resolution level includes one ResNet block. {The detailed hyperparameters of each U-Net subnetwork are summarized in \tabref{tab:unet_para}.}
The diffusion process uses $N = 40$ steps, and both DM and CM are trained for 800 epochs.  % 100000 steps x 8 / 1000

For the SIP scheme, the power allocation factor is selected as $\rho=0.3$. Pilot symbols are modulated using quadrature phase shift keying (QPSK), while data symbols employ 16-QAM. Low-density parity-check (LDPC) codes with a code rate of $r=490/1024$ are utilized. In the neural receiver, the number of interference cancellation iterations is set to $I=2$, and both the channel estimation and data detection modules use six ResNet blocks. 
{The receiver is trained for 300 epochs using the Adam optimizer with an initial learning rate of 0.001. During training, the SNR values are randomly selected within the range of $[-5, 0]$ dB.}
As a baseline, we include a conventional OP scheme using $N_{\rm p}$ pilot symbols, with linear minimum MSE (LMMSE)-based channel estimation and data detection.
% Metrics
To evaluate performance, the block error rate (BLER) is computed over 10,000 transmitted frames. The throughput is then calculated based on the BLER using the following equation:  % throughput per frame
\CheckRmv{
  \begin{equation}
    \text{Throughput} = K\times S\times \Omega\times Q \times r \times (1-\text{BLER}), 
  \end{equation}
}
where $\Omega$ represents the ratio of REs allocated for data symbols. Specifically, for the OP receiver, $\Omega = (S - N_{\rm p}) / S$, while for the SIP receiver, $\Omega = 1$.

%%%%%%%%%%%%%%%%%%%%%%%%%%%%%%%%%%%%%%%%%%%%%%%%%%%%%%%%%%

\subsection{Results and Analysis}
% {\rl Emphasize scenario-specific}
\CheckRmv{
  \begin{figure*}[t]
    \centering
    \includegraphics[width=6in]{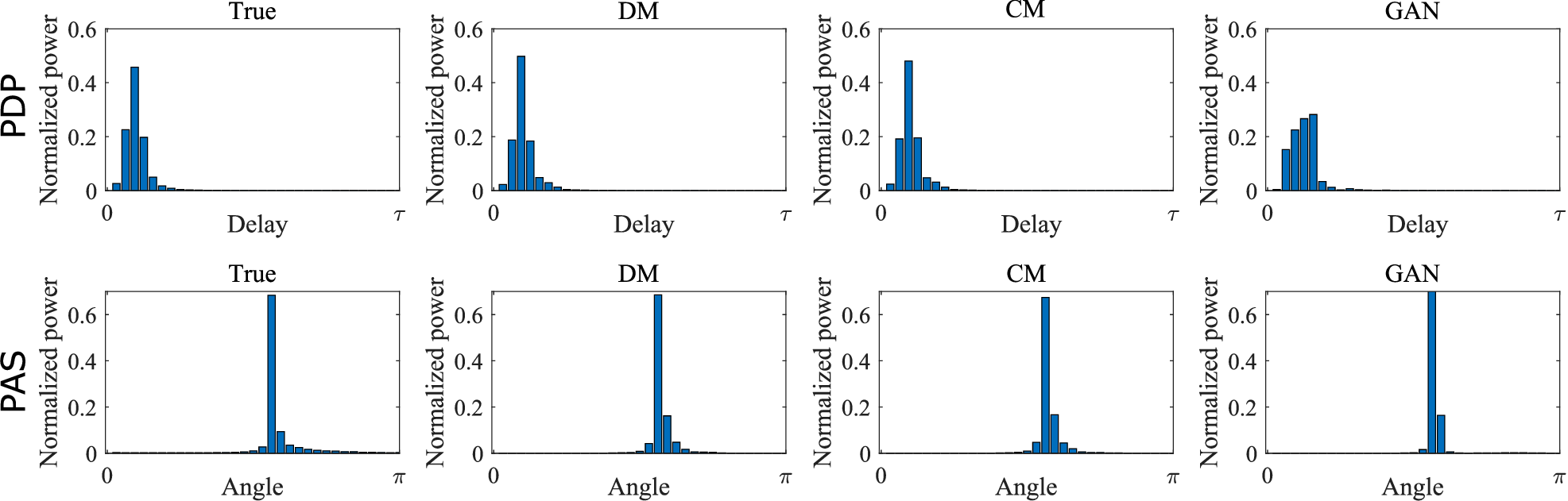}
    \caption{PDP and PAS of the channel samples from subregion R3.}
    \label{fig:pdp_pas}
  \end{figure*}
}

\CheckRmv{
  \begin{table}[t]
    \centering
    \begin{threeparttable}
    \caption{Wasserstein-2 Distance Between the Generated and True Channel Data across Different Subregions}
    \begin{tabular}{ll|ccccc} 
    \hline \hline
             &        &R1                  &R2                  &R3                  &R4       & R5         \\ 
    \hline
    \multirow{3}{*}{\begin{sideways} AD\end{sideways}} & DM  &\textbf{1.22e-2}             &9.86e-3             &7.27e-2              &6.20e-2                &\textbf{1.07e-2}  \\
    & CM  &1.33e-2             &\textbf{6.92e-3}            &\textbf{4.35e-2}              &\textbf{3.68e-2}                &1.45e-2  \\
    & GAN &{2.16e-1}    &4.07e-2             &{2.51e-1}             &{1.08e-1}       &{1.08e-1}  \\
    \hline
    \multirow{3}{*}{\begin{sideways} FT\end{sideways}} & DM   &\textbf{1.14e-4}      &{1.50e-4}  & \textbf{2.53e-5}             & \textbf{1.66e-5}    &{1.23e-4}  \\
    & CM   &{1.50e-4}           &\textbf{1.45e-4} &{7.44e-5}             &{5.52e-5}    &\textbf{1.17e-4}  \\
    & GAN   &{5.83e-4}           &{1.79e-4} &{1.06e-4}             &{3.25e-4}    &{1.29e-4}  \\
    \hline \hline
    \end{tabular}
    \label{tab:w2_dist}
    \begin{tablenotes}[para,flushleft]
      \footnotesize
      Note: {``AD'' refers to angular-delay domain. ``FT'' refers to frequency-time domain. The top-performing results are highlighted in \textbf{bold}.}
    \end{tablenotes}
  \end{threeparttable}
  \end{table}
}

\subsubsection{Evaluation of Generated Channel Samples}

In this subsection, we evaluate the quality of the channel samples generated by the proposed methods. We begin by comparing the statistical distributions of the true and synthetic channel samples, as illustrated in \figref{fig:pdp_pas}. For clarity, the comparison focuses on channel samples from subregion R3. As a baseline, we adopt the conditional GAN-based approach from \cite{brock2018large}. The power delay profile (PDP) and power angular spectrum (PAS) are calculated by averaging over 5000 channel samples. \figref{fig:pdp_pas} demonstrates that the PDP and PAS of the channels generated by the proposed DM and CM closely resemble those of the true channel samples used to train them. In contrast, the distributions of the samples generated by the GAN approach exhibit significant deviations from the true distribution. These results highlight the superior capability of the proposed methods in accurately modeling channel statistics and generating high-fidelity samples consistent with the underlying scenario.

% In addition to distributional similarity, \figref{fig:pdp_pas} also reports the generation latency for producing 5000 samples using each method. The proposed CM achieves a generation speed comparable to that of the GAN-based approach while providing approximately an 80-fold reduction in latency compared to the DM method, due to its single-step generation process. Importantly, this acceleration is achieved without compromising generation quality, demonstrating the practical efficiency and effectiveness of the CM architecture.  %Given that the generation quality virtually does not compromise, this acceleration is promising.  

To provide a quantitative assessment, \tabref{tab:w2_dist} presents the Wasserstein-2 distance between the power spectra of the generated and true channel samples. The computation follows the formulation in \cite[Eq.~(1)]{sengupta2023generative}. 
As shown in the table, the proposed DM- and CM-based approaches yield consistently smaller Wasserstein-2 distances compared to the GAN baseline across all subregions. This holds true in both the antenna-delay and frequency-time domains. These results confirm the superior ability of the proposed methods to align with the true channel distribution and reinforce their effectiveness in high-fidelity channel modeling.

\CheckRmv{
  \begin{table}[t]
    \centering
    {\begin{threeparttable}
    \caption{{KS Test P-values across Different UE Speed}}
    \begin{tabular}{l|ccc} 
    \hline \hline
             &DM              &CM                  &GAN                \\ 
    \hline
    24 km/h   &{2.17e-1}      &\textbf{3.02e-1}    &{2.59e-4}               \\
    300 km/h  &\textbf{2.43e-1}        &{1.42e-1}      &{4.28e-13}                \\
    \hline \hline
    \end{tabular}
    \label{tab:p-value}
    \begin{tablenotes}[para,flushleft]
      \footnotesize
      Note: {The top-performing results are highlighted in \textbf{bold}.}
    \end{tablenotes}
  \end{threeparttable}}
  \end{table}
}

{In addition to the Wasserstein-2 distance, we report the Kolmogorov-Smirnov (KS) test p-value to further quantify the distributional similarity between synthetic and true channel samples. A larger p-value indicates that the observed differences can be attributed to random variations, with 0.05 commonly adopted as the significance threshold \cite{wasserstein2016asa}.
Given the multi-dimensional nature of the channel samples, principal component analysis is first applied for dimensionality reduction, after which the KS test is conducted. 
The results for the frequency-time domain channel samples across different user mobility conditions are presented in \tabref{tab:p-value}. It can be observed that the GAN-based approach yields extremely small p-values, suggesting a clear mismatch with the true distribution. By contrast, the p-values of both the DM and CM exceed the significance level by a large margin, thereby confirming the effective distribution match achieved by these schemes.}

{Beyond generation quality evaluation, \tabref{tab:complexity} provides a comparison of DM, CM, and GAN-based approaches in terms of per-sample generation latency, parameter count, and FLOPs. All experiments were conducted on an NVIDIA RTX 4090 GPU to ensure consistency. As shown in the table, the proposed CM achieves a generation speed comparable to that of the GAN-based approach while providing approximately an 80-fold reduction in both sample generation latency and FLOPs compared to the DM method, due to its single-step generation mechanism. Importantly, this acceleration is achieved without compromising generation quality, demonstrating the practical efficiency and effectiveness of the CM architecture. In addition, the parameter counts of the proposed DM and CM remain comparable to those of the GAN. Taken together with their superior generation quality, these results underscore the strong potential of the proposed designs.}

\CheckRmv{
        \begin{table}[t]
            \centering
            {\begin{threeparttable}
            \caption{{Latency, Parameters and Complexity Comparison}}
                \begin{tabular}{l|ccc}
                \hline \hline 
                 & Latency [ms] & \#Params [M] & FLOPs [GFLOPs]  \\
                \hline \hline
                DM & 1033 & 144.86 & 16615  \\
                CM & 13 & 144.86 & 210.32  \\
                GAN & 12 & 156.76  & 21.13 \\
                \hline \hline
                \end{tabular}
            \label{tab:complexity}
            \begin{tablenotes}[para,flushleft]
            \footnotesize
            Note: {In the dual U-Net design of the DM/CM's denoising network $\boldsymbol{\epsilon}_{\boldsymbol{\theta}}$, the two subnetworks each account for approximately half of the total parameters and FLOPs. The parameter counts of the GAN's generator and discriminator are 70.31M and 86.45M, respectively.}
            \end{tablenotes}
            \end{threeparttable}}
        \end{table}
    }

\subsubsection{Evaluation of Receiver Performance}
To assess the effectiveness of the neural receiver under various channel augmentation strategies, we compare the following methods for constructing the training dataset.
\begin{itemize}  % todo: Ntrain, Ngen. In the following, the number of original available channel samples is $N_{\rm train}=1000$, the number of generated channel samples for augmentation is $N_{\rm gen}$.
  \item \textbf{True:} Uses ground-truth channel samples corresponding to the target user locations and speeds. 

  \item \textbf{GAN:} %Using the augmented dataset that combines the original available channel samples with the synthetic channel samples from the pre-trained GANs. 
  Augments the limited ground-truth samples with synthetic channel samples generated by a pre-trained conditional GAN.
  \item \textbf{DM (proposed):} Augments the limited ground-truth samples using synthetic channel data generated by a pre-trained conditional DM.
  \item \textbf{CM (proposed):} Augments the limited ground-truth samples using synthetic data generated by a pre-trained conditional CM.
  \item \textbf{Uncond. DM:} Augments the limited ground-truth samples using a pre-trained unconditional DM that does not incorporate scenario-specific conditioning information.
  \item \textbf{Noisy:} Augments the ground-truth samples by adding AWGN  to generate synthetic samples.
  {\item \textbf{Mixup:} Augments the ground-truth samples by linearly interpolating ground-truth channel tensors $\mathbf{h}_{\rm train}^{[i]}, \mathbf{h}_{\rm train}^{[j]}$ randomly drawn from $\mathcal{H}_{\rm train}$ to generate synthetic samples \cite{zhang2017mixup,zhang2020does}:
  \begin{equation}
    \hat{\mathbf{h}} = \lambda \mathbf{h}_{\rm train}^{[i]} + (1-\lambda)\mathbf{h}_{\rm train}^{[j]},
  \end{equation}
  where $\lambda \in [0, 1]$ is a mixing coefficient drawn from a Beta distribution, i.e., $\lambda\sim {\rm Beta}(\alpha, \alpha)$, $ \alpha>0$. Following common practice \cite{zhang2017mixup}, we set $\alpha=0.4$ in our experiments.}
\end{itemize}

\CheckRmv{
  \begin{figure*}[t]
    \centering
    \subfigure[BLER]{
      \includegraphics[width=3in]{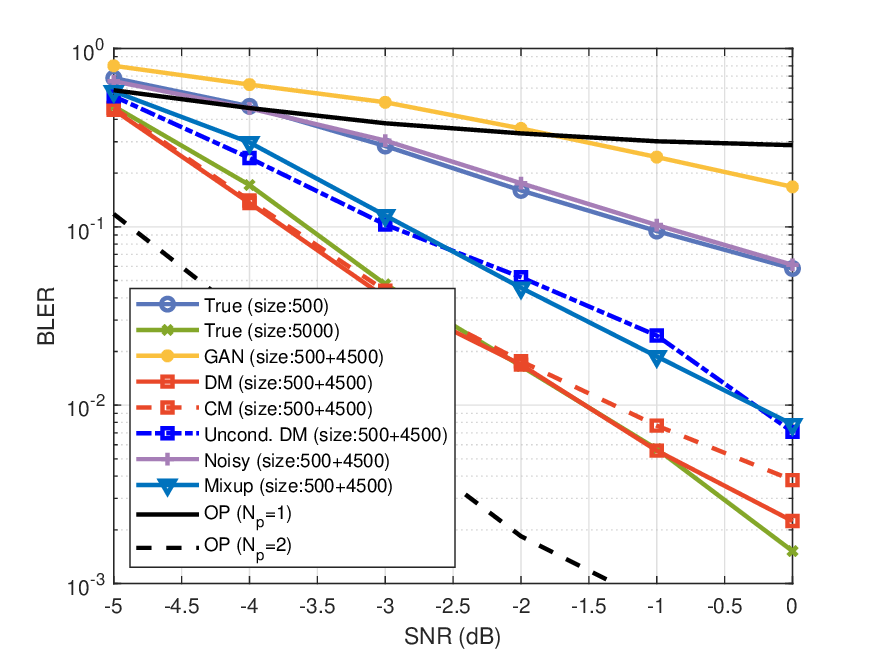}
      \label{fig:bler_24kmh}
    }
    \subfigure[Throughput]{
      \includegraphics[width=3in]{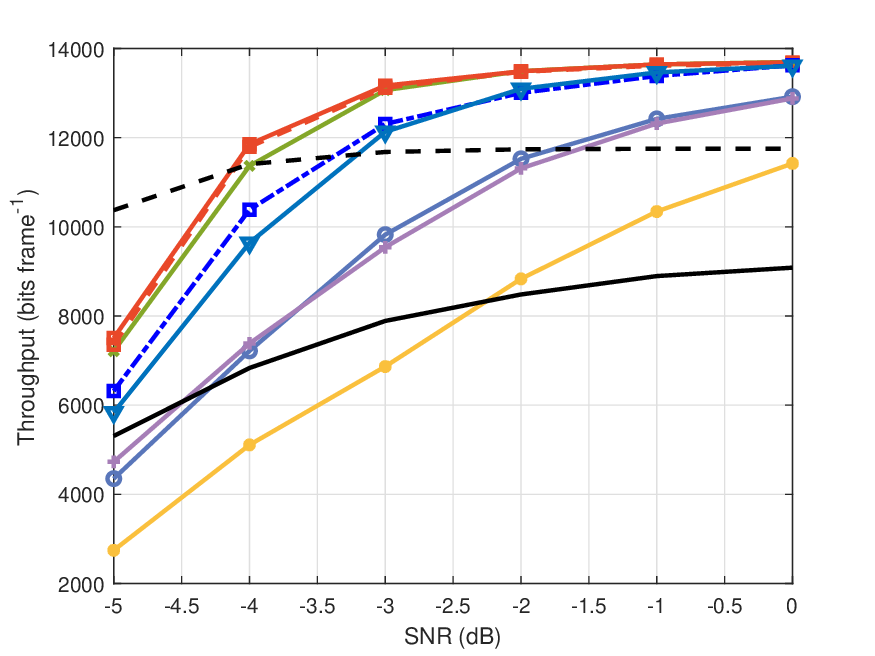}
      \label{fig:th_24kmh}
    }
    \caption{{BLER performance and throughput of various approaches for training the neural receiver when $v = 24$ km/h.}}
    \label{fig:24kmh}
  \end{figure*}
}

\CheckRmv{
  \begin{figure*}[t]
    \centering
    \subfigure[BLER]{
      \includegraphics[width=3in]{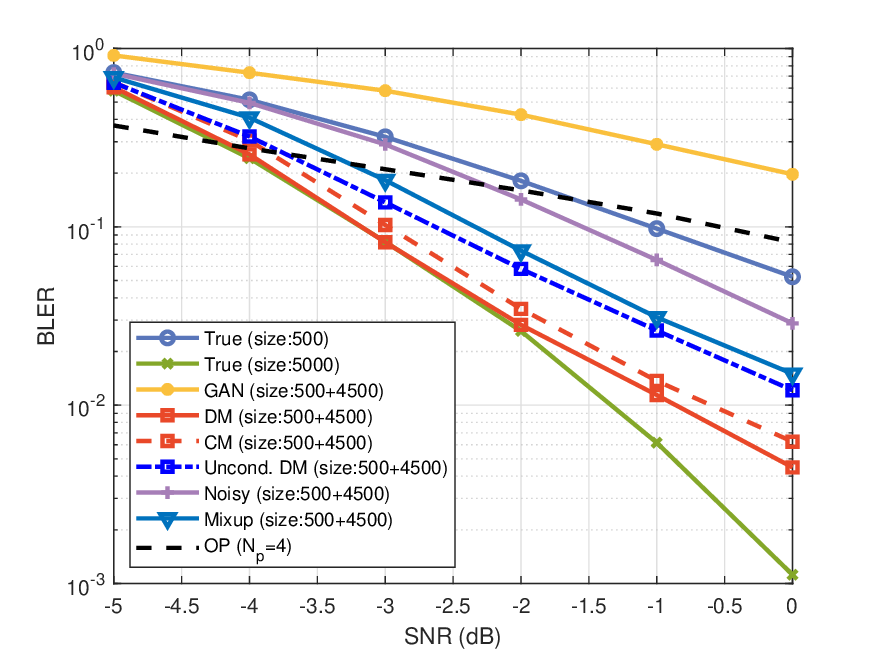}
      \label{fig:bler_300kmh}
    }
    \subfigure[Throughput]{
      \includegraphics[width=3in]{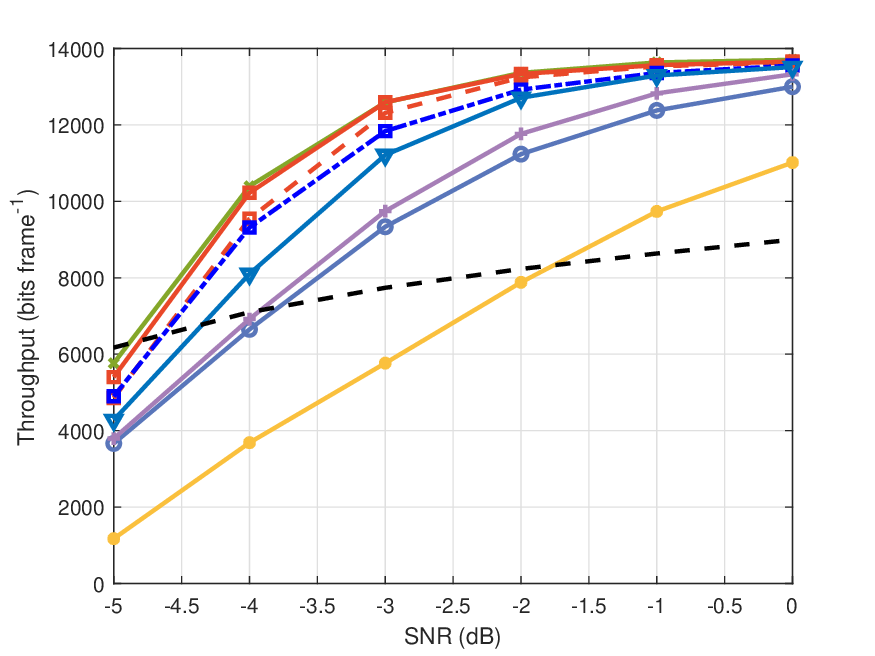}
      \label{fig:th_300kmh}
    }
    \caption{{BLER performance and throughput of various approaches for training the neural receiver when $v = 300$ km/h.}}
    \label{fig:300kmh}
  \end{figure*}
}

We first evaluate the performance of the SIP receiver under different user mobility conditions, focusing on both BLER and throughput, as depicted in Figs.~\ref{fig:24kmh} and \ref{fig:300kmh}. Accordingly, the proposed approaches enable user mobility-aware channel generation and augmentation by conditioning on labels $\mathbf{c}$ containing the corresponding user speed $v$. 

\figref{fig:bler_24kmh} presents the BLER results under a low speed of $v=24$ km/h. As described in the simulation setup, only a limited number of ground-truth channel samples are available at this speed, specifically $N_{\rm train}/2 = 500$ samples. To expand the training data, all augmentation methods, namely GAN, DM, CM, and AWGN, generate $N_{\rm gen} = 4500$ synthetic samples. For performance comparison, we also include a benchmark in which the receiver is trained using 5000 ground-truth channel samples.

As shown in the figure, receivers augmented with DM- and CM-generated samples achieve significant performance gains compared to the receiver trained solely on the limited original dataset. These improvements highlight the effectiveness of the proposed methods in alleviating the issue of data scarcity and enabling scenario-specific channel generation and augmentation. Moreover, the proposed methods perform comparably to the benchmark receiver trained with 5000 real samples and clearly outperform {the GAN-, AWGN-, and Mixup-based augmentation}. The inferior performance of the GAN-based method can be attributed to the limited diversity of its generated samples, which leads to overfitting during neural receiver training.

In addition, the enhanced SIP receiver exhibits a performance gap of approximately 1.5 to 2 dB compared to the OP receiver with $N_{\rm p} = 2$. However, as shown in \figref{fig:th_24kmh}, it provides a significant throughput advantage. This improvement results from the efficient use of all transmission resources enabled by the SIP scheme in conjunction with the proposed augmentation strategy. 

The BLER performance and throughput shown in \figref{fig:bler_300kmh} and \figref{fig:th_300kmh} demonstrate that the SIP scheme exhibits much more pronounced gains over the OP transmission under the high-mobility condition with $v=300$ km/h. The SIP receiver augmented by the proposed approach achieves similar BLER and throughput as those of the receiver trained using 5000 true channel samples, while also outperforming all other baseline methods. Notably, this receiver delivers a throughput improvement of approximately 51.8\% over the OP receiver with $N_{\rm p}=4$, highlighting the inability of limited pilot resources to track rapid channel variations in highly dynamic environments.

\CheckRmv{
  \begin{table}[t]
    \centering
    \begin{threeparttable}
    \caption{BLER Performance of the SIP Receiver across Different Subregions ($v = 24\;\mathrm{km/h}$, $\mathrm{SNR} =-3\;\mathrm{dB}$)}
    \setlength\tabcolsep{3pt}
    \begin{tabular}{l|ccccc} 
    \hline \hline
                     &R1                  &R2                  &R3                  &R4       & R5        \\ 
    \hline
    True 100 spec.  &9.16e-1             &7.66e-1              &8.90e-2              &3.64e-2                &4.25e-1  \\
    True 1000 spec. &\textbf{3.19e-2}    &\textbf{4.68e-3}     &\underline{3.12e-3}             &\textbf{4.76e-3}       &\underline{3.06e-2}  \\
    DM 1000 spec.   &\underline{3.38e-2} &\underline{5.00e-3}  &{3.72e-3}             &\underline{6.64e-3}    &\textbf{1.44e-2}  \\
    CM 1000 spec.   &{3.90e-2}           &{5.08e-3}            &\textbf{2.64e-3}              &8.88e-3                &4.20e-2  \\
    \hline
    True 5000       &1.06e-1             &1.63e-2             &4.04e-3             &1.10e-2                &8.75e-2  \\
    Uncond. DM 5000 & 2.45e-1             &1.72e-2             &3.24e-3             &1.62e-2           & 4.41e-2      \\
    \hline \hline
    \end{tabular}
    \label{tab:specific_aug}
    \begin{tablenotes}[para,flushleft]
      \footnotesize
      Note: {The top-performing results are highlighted in \textbf{bold}, with the second-best results \underline{underlined} for clarity.}
    \end{tablenotes}
  \end{threeparttable}
  \end{table}
}

% site-/location-specific generation: 1）分item来写  2）
\tabref{tab:specific_aug} presents the BLER of the SIP receiver across different subregions. %To show the site-specific augmentation capability of the proposed approach, we utilize the pre-trained conditional DM and CM to respectively generate 1000 samples for the five subregions using labels $\mathbf{c}$ containing the corresponding location coordinates. 
To enable site-specific channel generation and augmentation, the pre-trained conditional DM and CM are employed to generate synthetic samples based on label vectors $\mathbf{c}$, which encode the corresponding location coordinates.
For each subregion, $N_{\rm gen} = 1000$ synthetic samples are generated and %the generated samples 
combined with the 100 available ground-truth samples to construct an augmented dataset for training the receiver tailored to that specific subregion. 
% 和基线比
These approaches are denoted as ``DM 1000 spec.'' and ``CM 1000 spec.'', respectively.  As performance references, two baselines are considered: ``True 100'', which uses only 100 site-specific ground-truth samples, and ``True 1000'', which leverages 1000 ground-truth samples generated by QuaDRiGa. Results show that both proposed methods achieve comparable BLER as the ``True 1000 spec.'' baseline and significantly outperform the ``True 100 spec.'' baseline across all regions.
% 和general的比  For comparison,

In addition, two non-site-specific baselines are included for comparison: (1) ``True 5000'', trained on 5000 ground-truth samples drawn from all subregions, and (2) ``{Uncond. DM 5000}'', trained using 5000 synthetic samples generated by an unconditional DM. Since these models are not specialized for individual subregions, they adopt a shared receiver architecture across all subregions. The proposed site-specific methods exhibit substantially lower BLER than both baselines, underscoring the benefit of site-specific augmentation.

% \vspace{-0.5cm}

{\subsubsection{Ablation Study and Generalization Analysis}

\CheckRmv{
    \begin{figure}[t]
        \centering
        \includegraphics[width=3.0in]{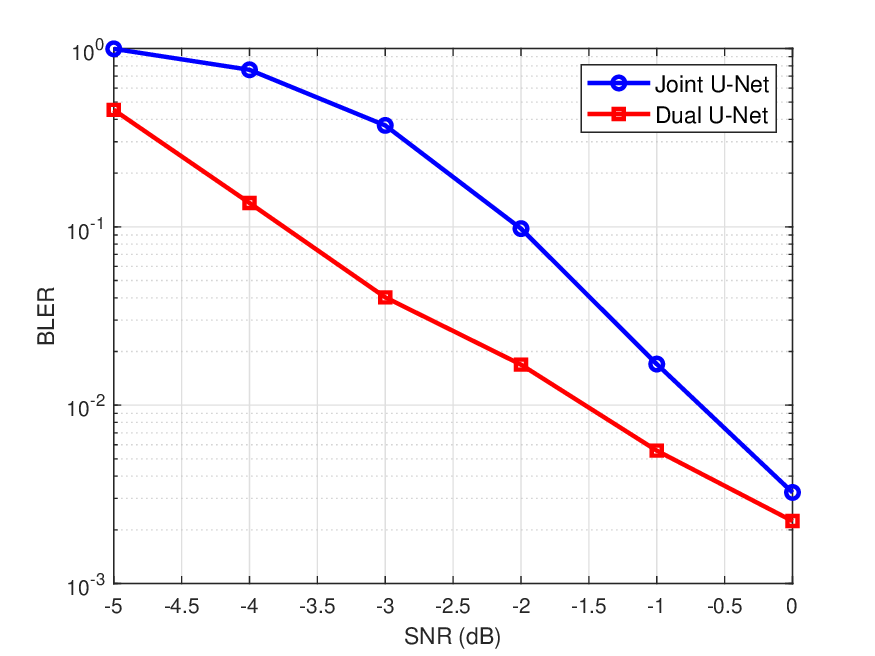}
        \caption{{Ablation study on the proposed dual U-Net architecture: BLER performance comparison between receivers augmented with the joint U-Net-based DM and the dual U-Net-based DM.}}
        \label{fig:ablation_joint}
    \end{figure}
}

This subsection presents the ablation study and generalization analysis of the proposed schemes.
First, to validate the effectiveness of the proposed dual U-Net architecture, we conduct an ablation study by training a single joint U-Net (with 162.89M parameters), which directly operates on the 2D ``image'' derived by flattening the angular-delay domain of the 3D CSI. The generated channel samples from this joint U-Net were then used to augment the SIP neural receiver, and the performance was compared with that of the receiver augmented by the proposed dual U-Net-based DM.
The BLER results under the mobility scenario of $v=24$ km/h, as presented in \figref{fig:ablation_joint}, clearly demonstrate that the proposed architecture significantly outperforms the single joint U-Net design, which fails to effectively capture the angular, delay, and temporal CSI features. These results justify our architectural choice.

\CheckRmv{
    \begin{figure}[t]
        \centering
        \includegraphics[width=3.0in]{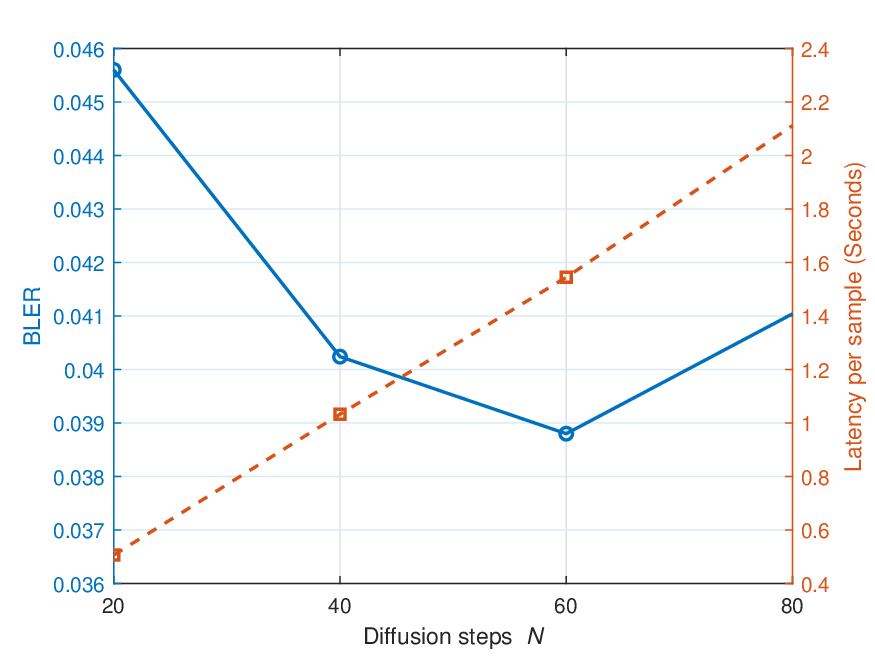}
        \caption{{Ablation study on the proposed DM-based approach's generation latency and BLER with respect to the number of diffusion steps $N$.}}
        \label{fig:ablation_n}
    \end{figure}
}

We then perform an ablation study to evaluate the impact of the diffusion-step parameter $N$ on the BLER performance and generation latency of the proposed DM-based augmentation approach, as presented in \figref{fig:ablation_n}. Specifically, BLER is evaluated under $\mathrm{SNR} =-3\;\mathrm{dB}$ and $v=24$ km/h with the augmentation setup of the DM-based approach identical to Fig.~5 of the manuscript, i.e., 500 ground-truth channel samples combined with 4500 synthetic samples for SIP neural receiver training. The results show that the BLER performance remains comparable when $N=40$, $N=60$, or $N=80$, indicating the robustness of the proposed DM-based approach with respect to the choice of $N$. Considering that the generation latency increases linearly with $N$, these results substantiate the choice of $N=40$ as the operating point of the proposed DM-based approach, as it provides a favorable balance between generation quality and efficiency.

\CheckRmv{
  \begin{figure}[t]
    \centering
    \includegraphics[width=3.0in]{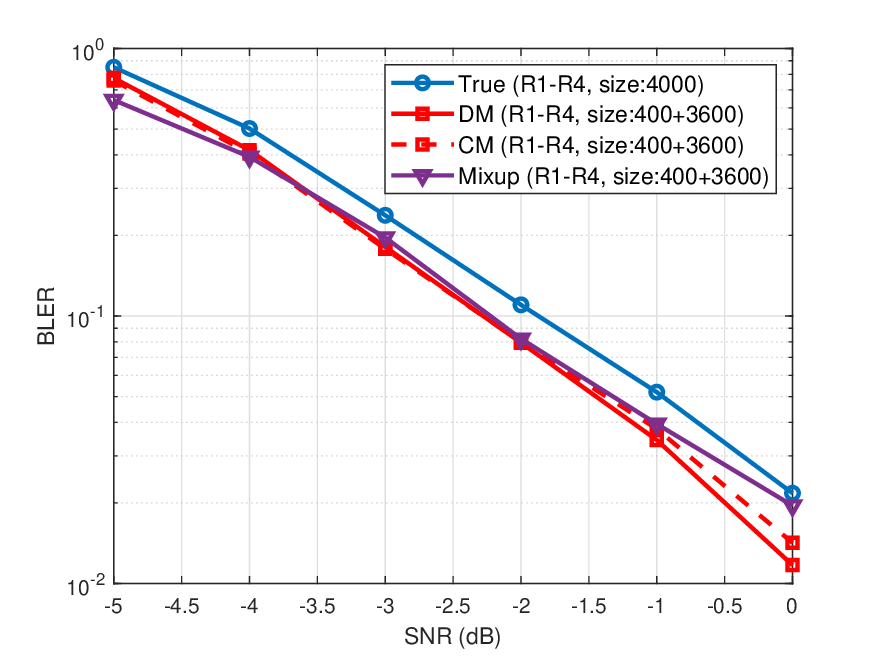}
    \caption{{Generalization study: BLER performance of various approaches for training the neural receiver when $v = 24$ km/h in subregion R5.}}
    \label{fig:generalization}
  \end{figure}
}

% We further evaluate the augmented SIP receiver in unseen scenarios.  
Finally, we perform a generalization study of the proposed approach by augmenting the SIP neural receiver with synthetic samples generated for subregions R1-R4 and testing on the unseen subregion R5. The corresponding BLER performance under $v=24$ km/h is reported in \figref{fig:generalization}. Specifically, the DM and CM each generate $N_{\rm gen} = 3600$ synthetic samples for R1-R4 (900 samples per region), which are combined with 400 ground-truth samples to form the augmented training dataset. For comparison, we also include the Mixup baseline and a benchmark receiver trained with 4000 ground-truth samples from R1-R4. As shown in the figure, Mixup, DM, and CM all outperform the benchmark. These results can be attributed to the increased diversity introduced by the synthetic samples, which enhances the generalization capability of the augmentation-based approaches. Furthermore, the proposed DM- and CM-based schemes provide further gains over Mixup, thereby confirming their effectiveness in enhancing robustness to scenario variations.}

% \vspace{-0.8cm}
  
%%%%%%%%%%%%%%%%%%%%%%%%%%%%%%%%%%%%%%%%%%%%%%%%%%%%%%%%%%
\section{Conclusion}   % 连reference 1.5 page
\label{sec:conclusion}
% \vspace{-0.3cm}

This paper proposes a scenario-specific channel generation method based on conditional DMs, designed to enhance neural receiver performance in SIP schemes. The proposed conditional DM is architecturally tailored to capture the complex relationship between channel characteristics and UE location and speed, thereby enabling the generation of high-fidelity channel data aligned with specific deployment environments.
To further improve generation efficiency, a consistency distillation technique is introduced, which significantly accelerates the sampling process. The generated synthetic channels effectively address the challenge of data scarcity in training neural receivers. 
Numerical results demonstrate that the proposed method mitigates the limitations imposed by limited training data and achieves scenario-specific performance that surpasses both general-purpose models trained across multiple scenarios and several existing data augmentation baselines.

%%%%%%%%%%%%%%%%%%%%%%%%%%%%%%%%%%%%%%%%%%%%%%%%%%%%%%%%%%
% \section*{ACKNOWLEDGEMENT}
% \label{ACKNOWLEDGEMENT}

% This work was supported by National Natural Science Foundation of China (No.~61821001) and Science and Tech-nology Key Project of Guangdong Province, China (2019B010157001).

%%%%%%%%%%%%%%%%%%%%%%%%%%%%%%%%%%%%%%%%%%%%%%%%%%%%%%%%%%
% \theendnotes

%%%%%%%%%%%%%%%%%%%%%%%%%%%%%%%%%%%%%%%%%%%%%%%%%%%%%%%%%%
\appendix

% \vspace{-0.8cm}
The one-step update rule of the Heun ODE solver, from $t_n$ to $t_{n-1}$ as presented in \algref{alg:sampling}, is derived in the following.
Define
\CheckRmv{
  \begin{equation}
    \boldsymbol{d}_{n} \triangleq \left. \frac{{\rm d} \hat{\mathbf{h}}_t}{{\rm d}t}\right|_{t=t_n} = \frac{\hat{\mathbf{h}}_{t_n} - \mathbf{d}_{\boldsymbol{\theta}}(\hat{\mathbf{h}}_{t_n}, t_n)}{t_n}, 
  \end{equation}
}
where $n \in \{1, \ldots, N\}$. Given Euler's first-order approximation
\CheckRmv{
  \begin{align}
    \left. \frac{{\rm d} \hat{\mathbf{h}}_t}{{\rm d}t}\right|_{t=t_n} &\approx \frac{\hat{\mathbf{h}}_{t_{n}}-\hat{\mathbf{h}}_{t_{n-1}}}{t_{n}-t_{n-1}},
    \label{eq:euler}
  \end{align}
}
we can derive the estimated data point at $t=t_{n-1}$ as 
\CheckRmv{
  \begin{equation}
    \hat{\mathbf{h}}^{\prime}_{t_{n-1}} \approx \hat{\mathbf{h}}_{t_n} + (t_{n-1} - t_n)\boldsymbol{d}_n. \label{eq:euler_tn-1}
  \end{equation}
}
Nonetheless, it is observed that Heun's second-order method achieves a desirable tradeoff between complexity and truncation error \cite{karras2022elucidating}. This method approximates the right-hand side of \eqref{eq:euler} using the derivative between $t_n$ and $t_{n-1}$:
\CheckRmv{
  \begin{align}
     \frac{\hat{\mathbf{h}}_{t_{n}}-\hat{\mathbf{h}}_{t_{n-1}}}{t_{n}-t_{n-1}} &\approx \frac{1}{2} \left. \frac{{\rm d} \hat{\mathbf{h}}_t}{{\rm d}t}\right|_{t=t_n} + \frac{1}{2} \left. \frac{{\rm d} \hat{\mathbf{h}}_t}{{\rm d}t}\right|_{t=t_{n-1}} \nonumber \\
     &= \frac{1}{2}\boldsymbol{d}_{n} + \frac{1}{2}\boldsymbol{d}_{n-1},
  \end{align}
}
where $\boldsymbol{d}_{n-1}$ is further approximated using $\hat{\mathbf{h}}^{\prime}_{t_{n-1}}$ in \eqref{eq:euler_tn-1} as 
\CheckRmv{
  \begin{equation}
    \boldsymbol{d}_{n-1} \approx \boldsymbol{d}_n^{\prime} = \frac{\hat{\mathbf{h}}^{\prime}_{t_{n-1}} - \mathbf{d}_{\boldsymbol{\theta}}(\hat{\mathbf{h}}^{\prime}_{t_{n-1}}, t_{n-1})}{t_{n-1}}.
  \end{equation}
}
Therefore, the estimation of $\hat{\mathbf{h}}_{t_{n-1}}$ can be formulated as
\CheckRmv{
  \begin{equation}
    \hat{\mathbf{h}}_{t_{n-1}} = \hat{\mathbf{h}}_{t_{n}} + (t_{n-1} - t_n)\left(\frac{1}{2}\boldsymbol{d}_{n} + \frac{1}{2}\boldsymbol{d}_{n}^{\prime}\right),
  \end{equation}
}
completing the derivation.

\bibliographystyle{IEEEtran}
\bibliography{IEEEabrv,myref}

\end{document}